\documentclass[final,5p,times,twocolumn]{elsarticle}

\usepackage{graphicx}
\usepackage{amssymb,amsmath}
\usepackage[stable]{footmisc}
\usepackage{lineno}
\usepackage{color}

\journal{Astroparticle Physics}

\begin{document}

\begin{frontmatter}

\title{Influence of the Geomagnetic Field on the IACT detection technique for possible sites of CTA observatories.}

\author[ul]{M.~Szanecki\corref{cor1}}
\ead{mitsza@uni.lodz.pl}

\author[mpik]{K.~Bernl\"ohr}
\ead{konrad.bernloehr@mpi-hd.mpg.de}

\author[ul]{D.~Sobczy\'nska}
\ead{ds@kfd2.phys.uni.lodz.pl}

\author[ul]{A.~Nied\'zwiecki}
\ead{niedzwiecki@uni.lodz.pl}

\author[ifae]{J.~Sitarek}
\ead{jsitarek@ifae.es}

\author[ul]{W.~Bednarek}
\ead{bednar@uni.lodz.pl}

\cortext[cor1]{Corresponding author}

\address[ul]{Department of Astrophysics, University of {\L}\'od\'z, Pomorska 149/153, PL--90--236 {\L}\'od\'z, Poland}
\address[mpik]{Max-Planck-Institut f\"ur Kernphysik, P.O.~Box 103980, D 69029 Heidelberg, Germany}
\address[ifae]{IFAE, Edifici Cn, Campus UAB, E-08193 Bellaterra, Spain}

\begin{abstract}
We investigate the influence of the geomagnetic field (GF) on the Imaging Air Cherenkov Telescope technique for two northern (Tenerife and San Pedro Martir) and three southern (Salta, Leoncito and Namibia (the H.E.S.S.-site)) site candidates for Cherenkov Telescope Array (CTA) observatories. 
We use the {\fontfamily{pcr}\selectfont CORSIKA} and {\fontfamily{pcr}\selectfont sim\_telarray} programs for Monte Carlo simulations of gamma ray showers, hadronic background and the telescope response. 
We focus here on gamma ray measurements in the low energy, sub-100 GeV, range. Therefore, we only consider the performance of arrays of several large telescopes. Neglecting the GF effect, we find (in agreement with previous studies) that such arrays have lower energy thresholds, and larger collection areas below 30 GeV, when located at higher altitudes. We point out, however, that in the considered ranges of altitudes and magnetic field intensities, 1800--3600 m a.s.l. and 0--40 ${\rm \mu T}$, respectively,  the GF effect has a similar magnitude to this altitude effect.
We provide the trigger-level performance parameters of the observatory affected by the GF effect, in particular the  collection areas, detection rates and the energy thresholds for all five locations, which information may be useful in the selection of sites for CTA. 
We also find simple scaling of these parameters with the magnetic field strength, which can be used to assess the magnitude of the GF effect for other sites; in this work we use them to estimate the performance parameters for five sites: South Africa-Beaufort West, USA-Yavapai Ranch, Namibia-Lalapanzi, Chile-La Silla and India-Hanle.
We roughly investigate the impact of the geophysical conditions on gamma/hadron separation procedures involving image shape and direction cuts. We note that the change of altitude has an opposite effect at the trigger and analysis levels, i.e.\ gains in triggering efficiency at higher altitudes are partially balanced by losses in the separation efficiency.
In turn, 
a stronger GF spoils both the shape and the direction discrimination of gamma rays, thus its effects at the trigger and analysis levels add up resulting in a significant
reduction of the observatory performance.  Overall, our results indicate that the local GF strength at a site can be equally important as its altitude for the low-energy performance of CTA.
\end{abstract}


\end{frontmatter}

\section{Introduction}
\label{sec:intro}
Imaging Air Cherenkov Telescopes (IACT) detect gamma rays using the Cherenkov images of their electromagnetic showers developing in the atmosphere. The IACT technique has rapidly advanced over the last 20 years (see, e.g., a review in  \citep{buckley08}) and, with the current generation of IACT instruments \citep{Aleksic11,Hofmann00,Veritas08}, it is now the most accurate and sensitive detection technique in the very high energy gamma ray astronomy.
The  Cherenkov Telescope Array (CTA), the next generation of IACT detectors, is expected to improve the sensitivity of present observatories by an order of magnitude, covering the energy range from a few tens of GeV to hundreds of TeV \citep{cta}.

In this paper we study the Geomagnetic Field (GF) effect, disturbing the IACT technique, which may set an inherent limit on the performance of CTA, especially at low energies. Charged particles in atmospheric showers are deflected by the Earth's magnetic field, which changes the geometry of the light pool\footnote{defined as the area on the ground with
nearly constant density of Cherenkov photons} (as directions of photons from e$^{+}$ and e$^{-}$ are deflected in opposite directions) and also leads to distortions of shower images.
This GF effect  was first pointed out in \citep{Cocconi54} and later  studied theoretically, e.g., by \citep{browning77,Elbert83,bowden92} and \citep{Commichau07,Commichau08} for the MAGIC telescope on La Palma. On the observational ground, the effect was confirmed   by several experiments, including the non-imaging Cherenkov telescope in Tien--Shan mountains \citep{Beisembayev99}, the Mark 6 telescope in Narrabri \citep{Chadwick99} and recently by the Pierre Auger Observatory \citep{PAO11}.

Although the GF effect is well understood, its magnitude at the candidate sites for CTA is rather uncertain. Previous detailed studies were done for Mark 6 \citep{Chadwick99} and MAGIC \citep{Commichau08} telescopes. However the obtained results were specific to the parameters (the telescope size, pixelization, etc.) of those instruments as well as to the local GF strength and direction at their locations.\\
\indent The GF effect affects the IACT measurements through two mechanisms. First, it stretches the light pool, decreasing (on average) the Cherenkov  photon density, thus reducing the probability of triggering the low energy showers with the IACT array.
In addition, the distortion of shower images spoils the gamma/hadron separation efficiency in the analysis of registered data and, furthermore, may affect the reconstruction of the gamma ray parameters such as the estimated energy and arrival direction. In this work we thoroughly investigate the first effect, i.e.\ the influence of GF on the probability of registering a gamma ray. 
We also quantify the changes of image parameters crucial for the gamma/hadron separation and the direction reconstruction, which allows us to discuss trends in the quality of the separation and reconstruction procedures for the changing magnitude of the GF effect. We emphasize, however, that our quantitative results for the triggering efficiency are more robust than these for the image-analysis level. The latter depend on a chosen analysis procedure, therefore, one can expect that the analysis-level performance  will be improved when optimized procedures are developed for a given site and telescope configuration. In contrary, the constraints on the triggering efficiency are intrinsic and will inevitably limit the observatory performance.\\
\indent At the trigger level, the GF effect is most pronounced in the low energy regime of a telescope; for gamma ray photons with energies well above the threshold, the Cherenkov pools have a sufficiently high density of photons even in strong GF.
In this work we rely on the CTA design concepts presented in \citep{cta}.
The proposed configurations include 3 classes of
telescopes, with large, medium and small size, planned to provide the optimum performance in different energy ranges. We focus here on the subarrays of several large telescopes, which are dedicated for high sensitivity observations below 100 GeV. 
Then, our results illustrate directly the impact of the GF effect on the energy threshold of the whole observatory, but qualitatively similar effects may be expected also in other classes of telescopes around their threshold energies.\\
\indent Apart from the local GF at an observatory location, the IACT performance at low energies may be significantly affected by its altitude  (cf.\ \citep{Aharonian01}). 
In this paper we consider five potentially interesting sites for CTA (cf.\ section 11 in \citep{cta}). 
We provide the site-specific information, however, we aim also to derive some more general properties. For each site we study different directions of observation corresponding to different strength of the magnetic field. Then, to disentangle effects due to the altitude and to the GF, for each site we analyse also the IACT performance with vanishing GF.
The considered sites span ranges of altitudes and the GF strength which are sufficient to derive some simple scaling formulas describing the dependence of performance parameters on both the GF strength and altitude. 
Thus obtained scaling laws allow us to estimate the magnitude of the GF effect 
for some other sites, which were indicated after the completion of our computations (see Sec.~\ref{sec:comp}).

\section{GF at the candidate sites}
\label{sec:sites}

Table \ref{tab:sites} gives the geophysical data for the sites considered in this paper.
We use the standard parametrization of the GF (see \citep{Campbell03}), 
$\vec{B}\equiv \left(H,0,Z \right)$, with the $x$-axis pointing to the local 
magnetic north, the $y$-axis pointing eastward and the $z$-axis oriented 
downwards.

Charged particles in the shower observed at the zenith angle $\theta$ and the azimuthal
angle $\phi$ are deflected by the  Lorentz force which is proportional to the component 
of $\vec{B}$ perpendicular to the observation direction:
\begin{align}
B_{\perp}\equiv|\vec{B}_{\perp}|=&\left[H^2\left(\cos^{2}(\theta)+\sin^{2}(\theta)\sin^{2}(\phi)\right)\right.\nonumber\\
&\left.-HZ\sin(2\theta)\cos(\phi)+Z^2\sin^{2}(\theta)\right]^{\frac{1}{2}}.
\label{Eq:Magnet_Perp_1}
\end{align}
Note that in our definition the azimuthal angle, $\phi$, is measured with respect 
to the local magnetic south; Table \ref{tab:sites} gives the magnetic declination angle $\delta$ (i.e.\ the angle between the local magnetic north and the geographic north directions).
The magnetic inclination angle $i$ in Table \ref{tab:sites} is defined as an acute angle between  $\vec{B}$ and the vertical direction. For convenience $i$ is measured with respect to positive and negative $z$-axis direction for northern and southern sites, respectively; we use a non-standard definition of $i$ in order to make direction-dependent effects in the observatory performance easier to assess (thus defined $i$ corresponds to the observation angle $\theta$).

\begin{table}[t]
\begin{center}
 \begin{tabular}{|c|c|c|c|c|c|c|c|}
  \hline
  \multicolumn{3}{|c|}{Site} & {\bf A-S} &{\bf A-L} & {\bf M} & {\bf S} & { \bf N} \\
  \hline
  \multicolumn{8}{|c|}{Localization}\\
  \hline
  \multicolumn{3}{|l|}{Latitude [$^{\circ}$]} & -24.1 & -31.1 & 31 & 28.3 & -23.4\\
  \multicolumn{3}{|l|}{Longitude [$^{\circ}$]} & -66.2 & -69.3 & -115.5 & -16.5 & 16.5\\
  \hline
  \multicolumn{3}{|l|}{Height [m a.s.l.]} & 3600 & 2660 & 2400 & 2200 & 1800\\
  \hline
  \multicolumn{8}{|c|}{Local magnetic field $\vec{B}=(H,0,Z)$}\\
  \hline
  \multicolumn{3}{|c|}{$H$ [$\mu$T]}& 21.1 & 20.1 & 25.3 & 30.6 & 12.1\\
  \multicolumn{3}{|c|}{$Z$ [$\mu$T] (+ Down)}& -8.8 & -12.2 & 38.4 & 23.2 & -25.5\\
  \hline
  \multicolumn{3}{|l|}{Declination $\delta$ [$^{\circ}$]}& -5.9 & 0.2 & 11.4 & -5.9 & -13.6\\
  \multicolumn{3}{|l|}{Inclination $i$ [$^{\circ}$]}& 67 & 59 & 33 & 53 & 25\\
  \hline
  \multicolumn{8}{|c|}{$B_{\perp}(\theta,\phi)$ [$\mu$T]}\\
  \hline
  \multicolumn{1}{|c|}{$\theta$=$30^{\circ}$}&\multicolumn{2}{|l|}{$\phi$=$0^{\circ}$}& 22.7 & 23.5 & 2.7 & 14.9 & 23.8\\
  \multicolumn{1}{|c|}{$\theta$=$30^{\circ}$}&\multicolumn{2}{|l|}{$\phi$=$180^{\circ}$}& 13.9 & 11.3 & 41.1 & 38.1 & 2.3\\
  \hline
 \end{tabular}
\end{center}
\caption{The geophysical data for Argentina-Salta ({\bf A-S}), Argentina-Leoncito ({\bf A-L}), M\`exico-San Pedro Martir ({\bf M}), Spain-Tenerife ({\bf S}) and Namibia-H.E.S.S. ({\bf N}) sites. The values of $H$, $Z$, $\delta$, $i$ and $B_{\perp}$ are obtained from the data given by {\it National Geophysical Data Center} at {\it www.ngdc.noaa.gov/geomag}; for each site we take the GF parameters for the altitude of 10 km a.s.l. The positive value of the declination angle $\delta$ is measured from North towards East directions.}
 \label{tab:sites}
\end{table}

\begin{figure}[t!]
\begin{center}
\includegraphics[trim = 4.5mm 0mm 0mm 2mm, clip,scale=0.45]{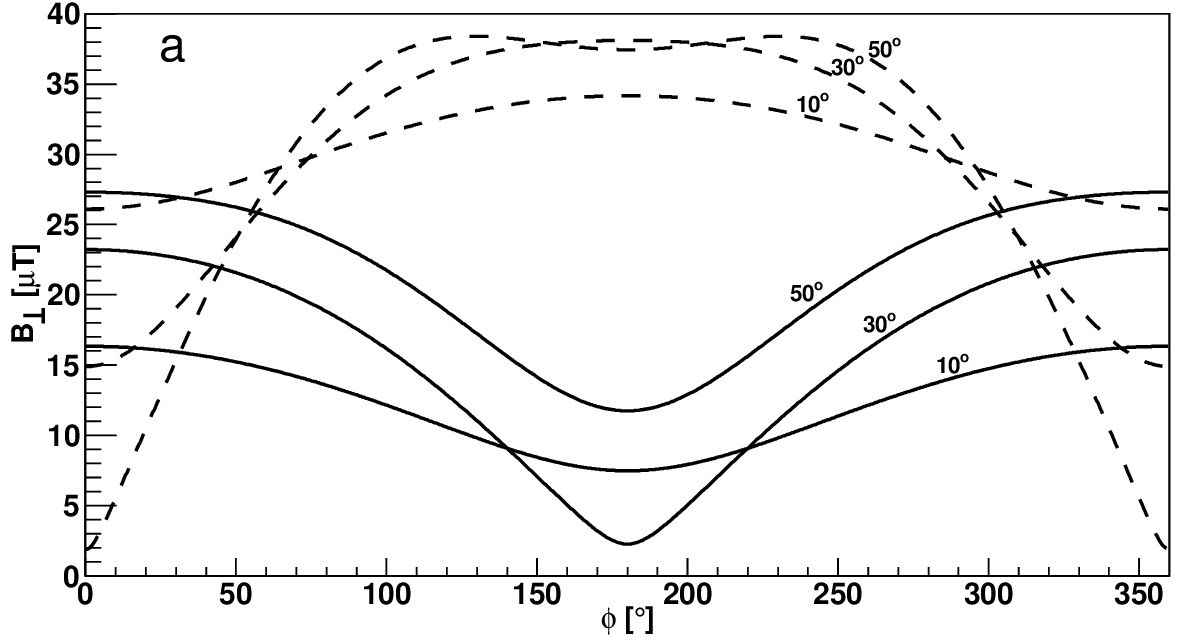}
\includegraphics[trim = 4.5mm 0mm 0mm 2mm, clip,scale=0.45]{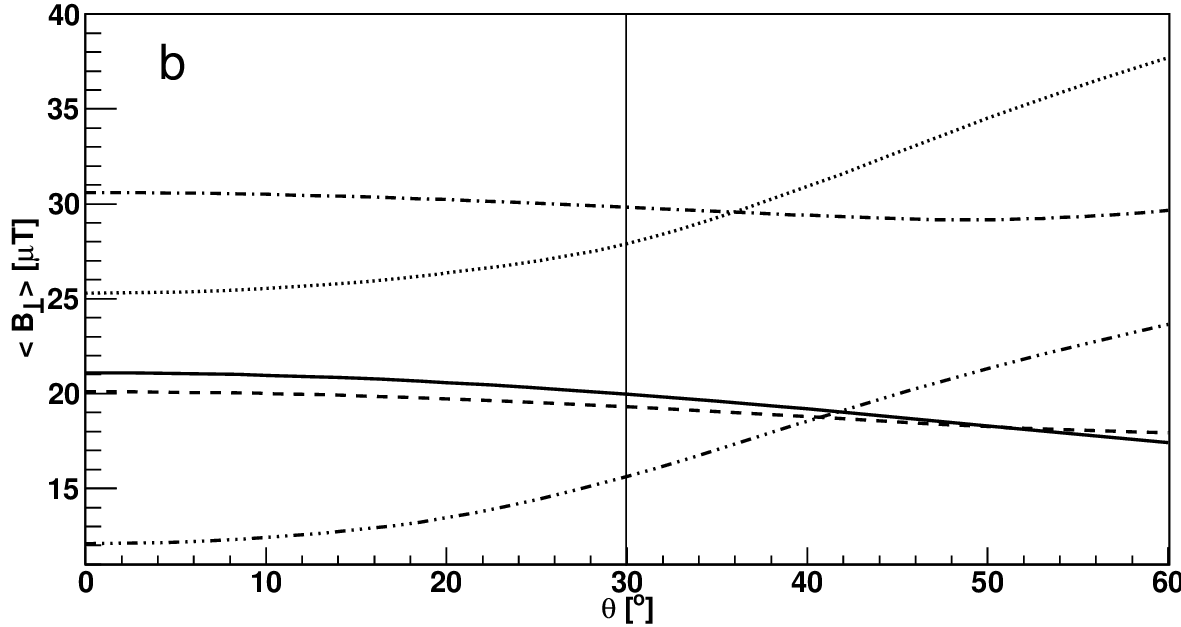}
\caption{Dependence of the transverse component of GF, $B_{\perp}$, on the observation direction. (a) Dependence on $\phi$ for fixed $\theta$=10$^\circ$, 30$^\circ$, 50$^\circ$
for the H.E.S.S. (the solid lines) and Tenerife (the dashed lines) sites; each line is labeled by the corresponding value of the Zenith angle, $\theta$. (b) The normal component of GF averaged over $\phi$, $\left<B_{\perp}\right>$, as a function of the zenith angle for: Argentina-Salta (the solid line), Argentina-Leoncito (the dashed line), Namibia (the triple-dot-dashed line), San Pedro Martir (the dotted line) and Tenerife (the dot-dashed line).}
\label{fig:b}
\end{center}
\end{figure}

Note that, in general, the GF is weaker in the southern sites (considered here) than in the northern ones.
The Argentinian sites are located in the region of the South Atlantic 
Anomaly (SAA). The Namibian site is located at a rather large distance from the minimum of SAA
but still it has a rather weak GF. The strongest GF occurs at the Mexican site. 

Fig.\ \ref{fig:b} illustrates the dependence of the transverse GF on the observation direction.
The change of the sign of the $Z$ component between the southern and northern hemisphere results in an 
opposite dependence on the azimuthal angle in southern and northern sites. In the latter, the normal component of $\vec{B}$ has smaller values--and, hence, the GF effect is weaker--for observation in  southward directions (dashed curves in Fig.\ \ref{fig:b}a). For southern sites, the GF effect is weaker for observations in northward directions (solid curves in Fig.\ \ref{fig:b}a). 

Finally, we discuss the dependence on the zenith angle. As the absorption in
the atmosphere increases with zenith angle, typically sources are not
observed by IACT instruments with zenith angle larger than 50-$60^{\circ}$. Large zenith angles are not suitable for observation at low energies, as the density of Cherenkov photons, for a given gamma ray energy, decreases significantly for $\theta > 30^{\circ}$.
For example, the energy threshold increases by 30\% with the increase of $\theta$ from  $0^{\circ}$ to $30^{\circ}$, but then it increases by a factor of three with the increase of  $\theta$ to $50^{\circ}$, according to simulations performed for the MAGIC I telescope \cite{Firpo06} and the HESS telescope array \cite{Aharonian06}.
 For our simulations we  choose $\theta = 30^{\circ}$, which allows to study large ranges of the strength of the transverse GF by changing  $\phi$ and moreover is a critical for observations in low energy range.

The dependence on the zenith angle at a given site results from the relation between the horizontal, $H$, and the vertical, $Z$, intensities. 
At the Namibian and Mexican sites, the $H$ component is larger than the $Z$ component.
At these sites, the inclination angle
of $\vec{B}$ is $i \approx 30^{\circ}$, see Table \ref{tab:sites}. For both Argentinian sites as well as for Tenerife, the horizontal intensity is much larger than the vertical intensity and the inclination is $i \approx 60^{\circ}$.

A shower developing along the magnetic field line is not broadened by the GF. In the Namibian and Mexican sites such showers are observed at $\phi = 180^{\circ}$ and $0^{\circ}$, respectively, and $\theta \approx 30^{\circ}$. 
At these sites, the change of $\phi$ for a fixed $\theta = 30^{\circ}$ yields a large amplitude of variation of $B_{\perp}$, as shown, for the Namibian site, by the solid curves in Fig.\ \ref{fig:b}a.
At the remaining three sites, the amplitude of variation of $B_{\perp}$ increases with increasing  $\theta$, with the maximum variation occurring for $\theta \approx 60^{\circ}$, as shown by the dashed curves in Fig.\ \ref{fig:b}a; $\theta=30^{\circ}$ corresponds to moderate changes at these sites. 

Fig.\ \ref{fig:b}b shows the transverse GF averaged over $\phi$, $\left<B_{\perp}\right>$, as a function of $\theta$.
As we can see, for Tenerife and the Argentinian sites (where $H > |Z|$), the average $\left<B_{\perp}(\theta)\right>$ has approximately the same value at all $\theta<60^{\circ}$. At the Namibian and Mexican sites (where $H < |Z|$), the average $\left<B_{\perp}(\theta)\right>$ increases significantly for $\theta>30^{\circ}$, nevertheless, $\left<B_{\perp}(30^{\circ})\right>$ is representative for zenith angles, $\theta \le 30^{\circ}$, relevant for observations in the low energy range.
 We refer to these properties in Sect.\ \ref{sec:comp}.

\section{MC simulations}
\label{sec:mc}

\begin{table}[t]
\begin{center}
 \begin{tabular}{|c||c||}
  \hline
  {\fontfamily{pcr}\selectfont Sim\_telarray} input & Input value\\
  \cline{1-2}
  \hline \hline
  Telescope type & LST (Type 1)\\
  \hline
  Dish diameter & ${\rm d}=24 \;{\rm m}$\\
  \hline
  Focal length/diameter & ${\rm f/d}=1.3$\\
 \hline
  Camera  Field of view &  ${\rm FoV}=5^{\circ}$\\
  \hline
  Pixel size & $0.09^{\circ}$\\
  \hline
  Photomultipliers & bi-alkali\\
  quantum efficiency &  ${\rm QE}_{\rm peak}=25.7\%$\\
  \hline
  Telescope trigger & Min.\ 4 pe in each of\\
  threshold level & 3 neighboring pixels\\
 \hline
  Min.\ trigger multiplicity & 2 telescopes \\
  \hline
   \end{tabular}
\end{center}
\caption{Parameters assumed in {\fontfamily{pcr}\selectfont sim\_telarray} for our simulations.}
 \label{tab:mc}
\end{table}

We use CTA Monte Carlo tools to simulate  the gamma ray, proton shower development and the telescope response and, then, we apply image-analysis procedures to the simulated data.

The EAS (Extensive Air Shower) simulations are performed using the {\fontfamily{pcr}\selectfont CORSIKA 6.98} code adapted for CTA \citep{Heck98,Bernlohr08}; the code includes the description of the influence of GF on EAS.

At each site we use the same, tropical atmospheric profile (see e.g.\ \cite{Bernlohr00}), which is appropriate for all sites considered in our work (the expected differences in the density of the Cherenkov light, resulting from differences in the air density profiles, do not exceed a few per cent). Concerning the atmospheric transmission, at each site we use the transmission coefficients calculated with MODTRAN \citep{Kneizys96} for a desert region with a very clear atmosphere (with the aerosol profile similar to that of the 'navy maritime atmosphere' discussed in \cite{Bernlohr00}); at each site we use the transmission coefficients appropriate for its altitude. Obviously, the atmospheric transmission may be expected to vary from one site to another, and even at a given site large seasonal changes may be expected, resulting in changes of the density of the Cherenkov light by over an order of magnitude. This should be an important factor for site selection, however, during the writing of this paper even a preliminary outcome of the atmospheric monitoring of the candidate sites, in terms of extinction, were not known (as noted in \cite{Bernlohr00} the model used here reproduces very well the extinction measured on La Palma during clear conditions, so it should be valid also for Tenerife).

We simulate showers induced by primary gamma rays from a point-like source and by protons arriving from a cone with an opening full angle of $14^{\circ}$. The basic parameters for simulations of gamma rays are:

\noindent
(i) energies between 3 GeV and 1 TeV generated from a power-law spectrum with the photon spectral index $\Gamma = 2$. 
The value of $\Gamma = 2$ is used to achieve a sufficient number of simulated showers in the whole energy range. However, we use event weights to re-weight the primary spectrum to $\Gamma=2.6$ or $\Gamma=3.5$, which slopes are more typical for gamma ray sources. Then, most of results presented in this work correspond to $\Gamma=2.6$, but we check the dependence on the value of $\Gamma$ by comparing them with the results for $\Gamma=3.5$.

\noindent
(ii) a fixed Zenith angle $\theta$=$30^{\circ}$; 

\noindent
(iii) the impact points generated up to $D=800$ m from the center of the array; we have checked that this maximum impact distance is sufficient to properly describe the performance of considered IACT arrays in the assumed range of energies,

\noindent
At each site we consider the azimuthal angles $\phi=0^{\circ}$ and $180^{\circ}$, corresponding to the minimum and maximum value of $B_{\perp}$. We also make a more detailed study for one specific site, in Namibia, by considering seven azimuthal angles between $\phi=0^{\circ}$ and $180^{\circ}$ with uniform steps of $30^{\circ}$. For each site we consider also the case of vanishing GF, to illustrate effects due to the change of the altitude. The same number of $5\times10^{6}$ events were simulated in our analysis in each case, in particular, for each location and each direction of observation.

For protons we use:

\noindent
(i) energies between 10 GeV and 3 TeV generated from a power-law spectrum with index $\Gamma = 2$ re-weighted then to $\Gamma=2.73$;

\noindent
(ii) Zenith angle $\theta$=$30^{\circ}$;

\noindent
(iii) view cone of  $14^{\circ}$ (full angle);

\noindent
(iv) the impact points generated up to $D=1500$ m from the center of the array.

\noindent
Due to small fraction of triggering events, for hadronic background we need at least one order of magnitude more generated events than for gamma rays. Thus, for protons we consider only three sites with neglected GF, i.e. Argentina-Salta, Tenerife and Namibia and  one with GF (Tenerife) with angles $\phi=0^{\circ}$ and $180^{\circ}$, corresponding to the minimum and maximum value of $B_{\perp}$. For each case  $2\times10^{8}$ protonic events were simulated. 

\begin{figure}[t]
\begin{center}
\includegraphics[trim = 20mm 40mm 0mm 1mm, clip,scale=0.4]{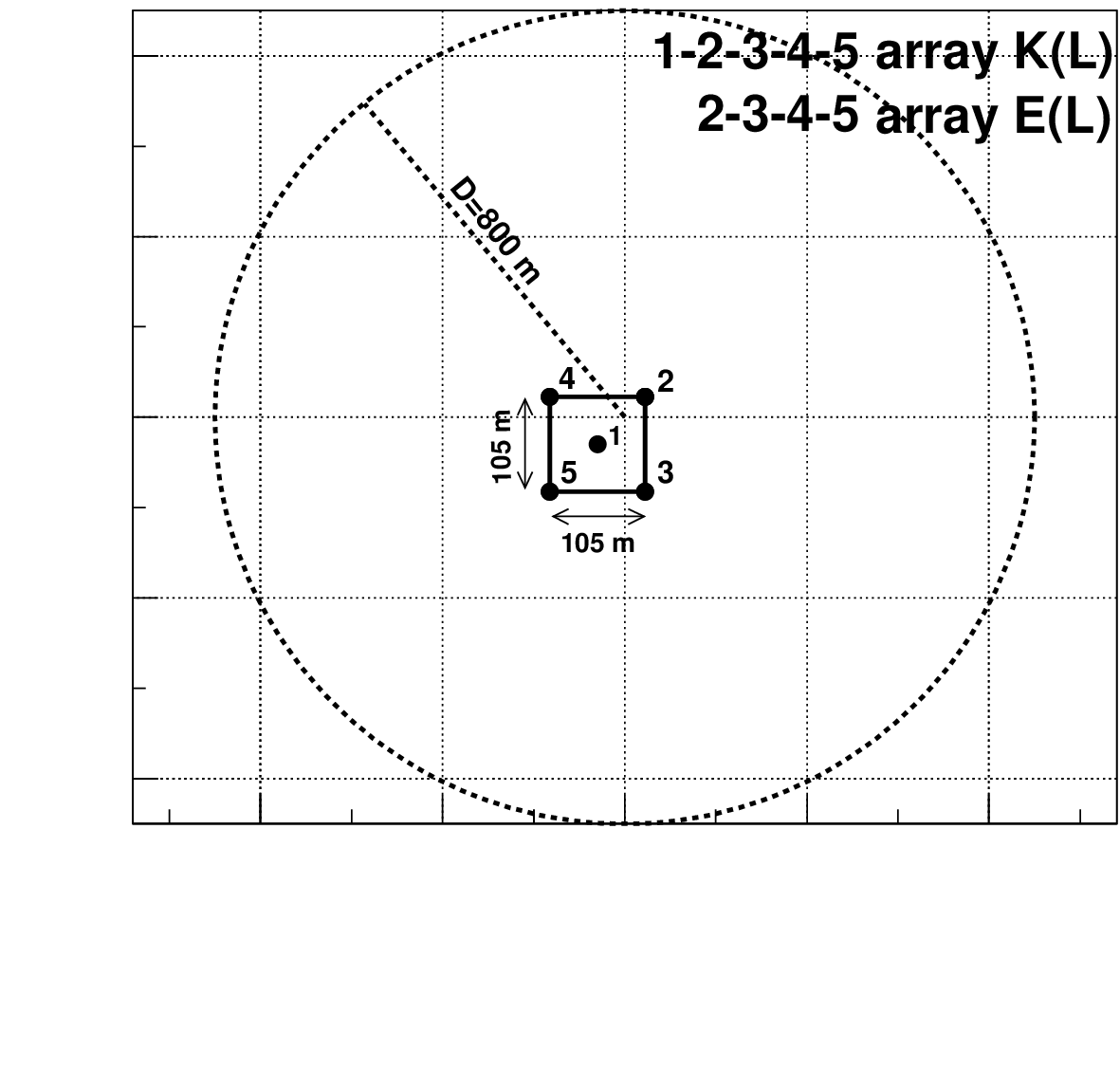}
\caption{Geometric layout of telescopes used in our simulations of gamma rays. The circle shows the simulated area defined by the maximum impact parameter, $D$.}
\label{fig:arrays}
\end{center}
\end{figure}

For the simulations of the telescope response we use the CTA {\fontfamily{pcr}\selectfont sim\_telarray} software \citep{Bernlohr08, BernlohrCTA08} with the default parameter set of the telescope and camera systems assumed in the {\em production-1}: first CTA MC mass production (see, Fig.~18 and Chapt.~8 of \citep{cta} or Chapt.\ 6 of \cite{Bernlohr12}). The {\em production-1} parameters crucial for our simulations are given in Table \ref{tab:mc}. Obviously, the performance parameters discussed in the next section depend on these assumptions. However, having established a particular set of the telescope and camera parameters, we can compare the performance of an IACT observatory at various GF strengths or altitudes, as changes of these assumptions should lead to systematic improvement or worsening of the IACT performance. As an example,
we have investigated various trigger configurations, with different number of triggered telescopes, $N$, required to register an event. However, we found that there are no noticeable differences between our results for different $N$, especially concerning the GF effect, apart from the obvious property of systematically higher energy thresholds and smaller detection rates for systems with higher $N$. Then, in this paper we present only our results for $N=2$.

We consider two telescope layouts presented in Fig.~\ref{fig:arrays}, namely arrays E(L) and K(L) with 4 and 5 Large Size Telescopes (LST), respectively.  We emphasize that arrays E(L) and K(L), studied in this work, refer to full CTA arrays E and K composed of only LSTs. Full arrays E and K, which were commonly used in design studies for CTA, contain large number of medium and small telescopes, additionally.
Arrays E(L) and K(L) (see Fig.~\ref{fig:arrays}) are slightly offset from the center of the simulated area, as they are part of a full, geometrically optimised CTA layout. Nevertheless, final results are not modified as the simulated area is large enough to enclose all the triggered events.
All the results presented in Sect.\ \ref{sec:results} correspond to array E(L); the systematic changes between arrays E(L) and K(L) are briefly summarized in Sect.\ \ref{sec:4.1}. 
For the sake of simplicity we have assumed the same night sky background\footnote{The rate of the accidental stereo events due to NSB is of the order of 100 Hz.} for each site; we have also assumed
the same array spacing equal to 105 m (see Fig.~\ref{fig:arrays}), we briefly comment on this assumption in Section 4.

\begin{figure}[t!]
\begin{center}
\includegraphics[trim = 8mm 3mm 0mm 2mm, clip, scale=0.45]{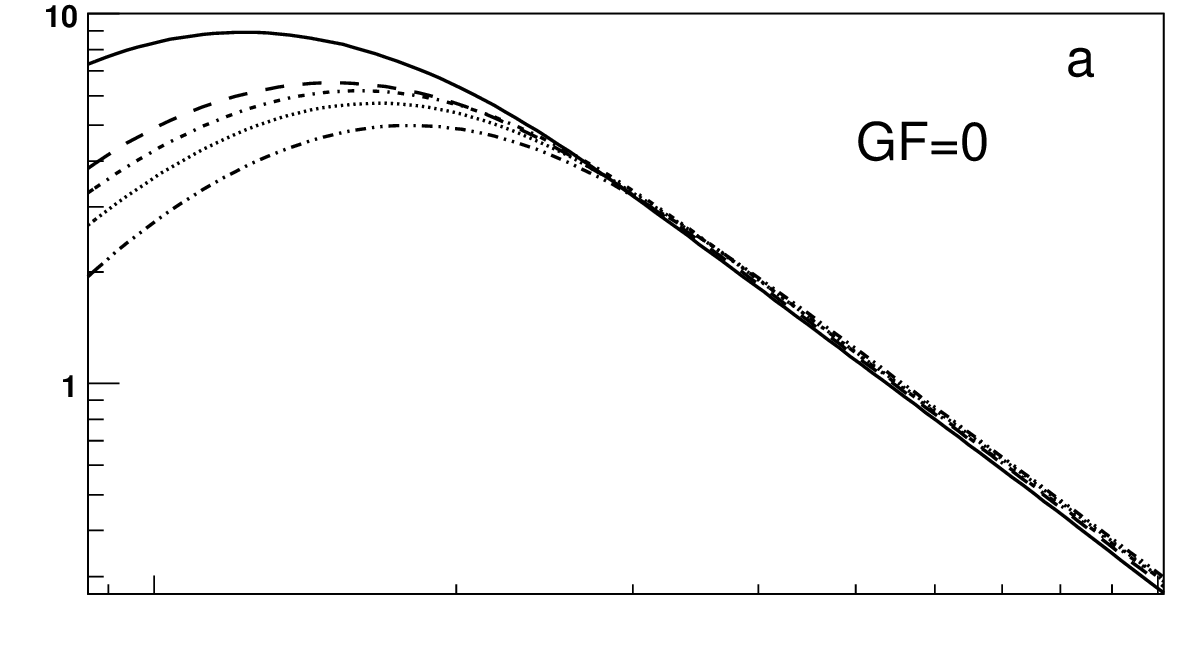}
\includegraphics[trim = 8mm 3mm 0mm 2mm, clip, scale=0.45]{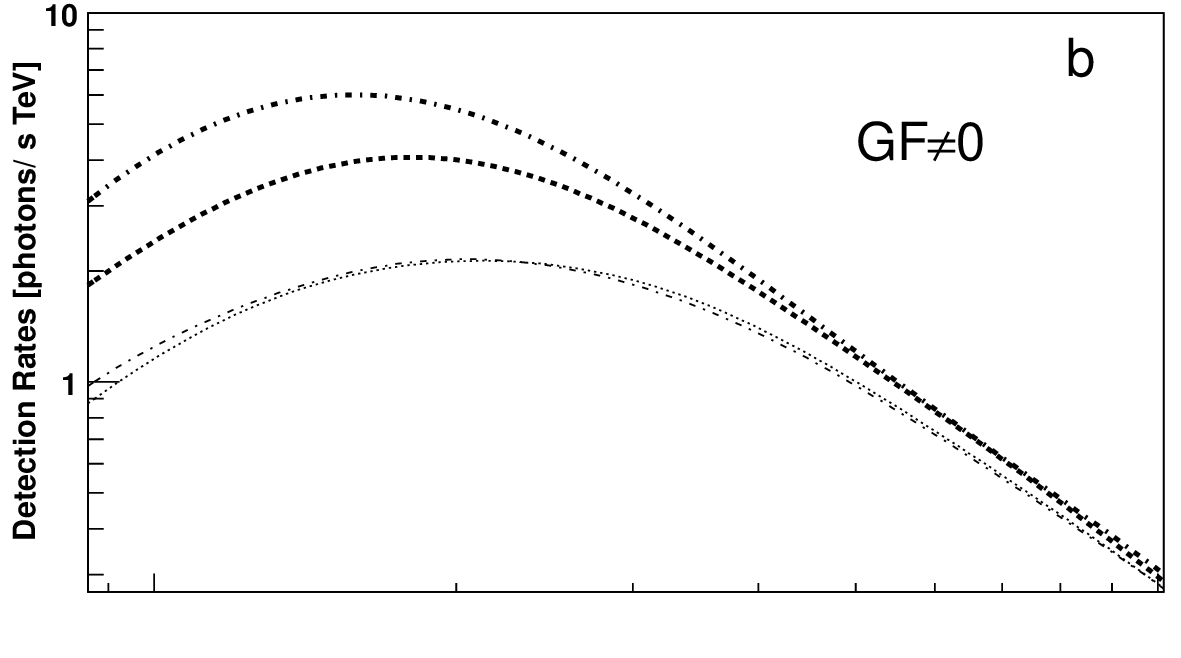}
\includegraphics[trim = 8mm 3mm 0mm 2mm, clip, scale=0.45]{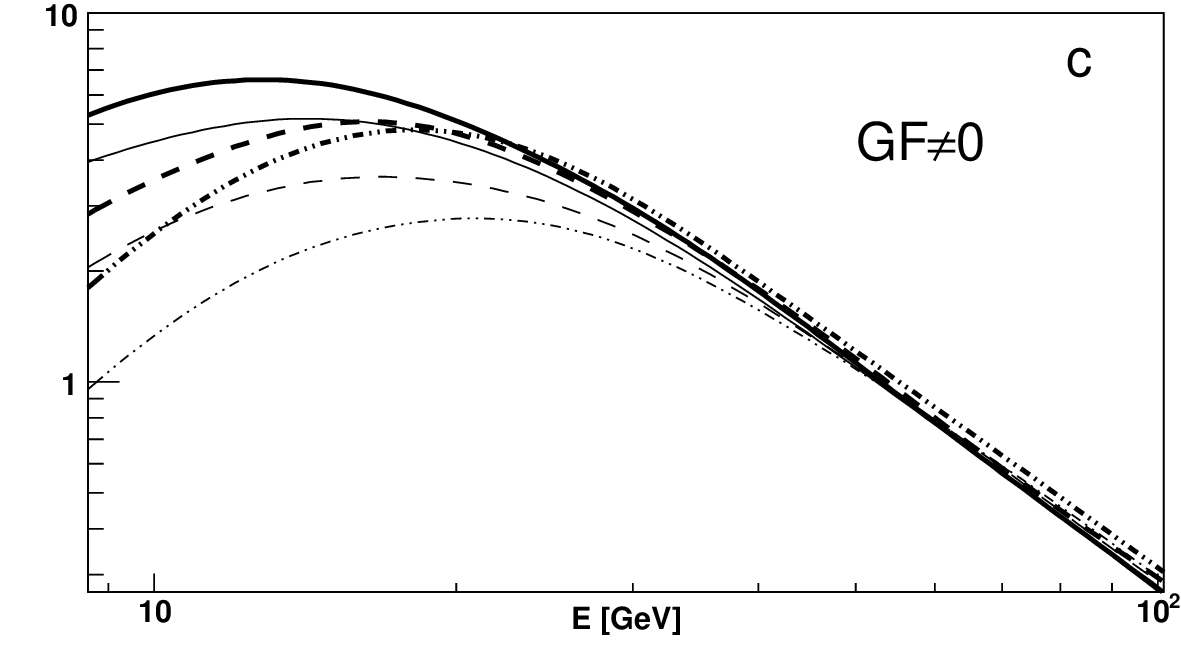}
\caption{Differential trigger rates (smoothed histograms) for Argentina-Salta (the solid line), Argentina-Leoncito (the dashed line), San Pedro Martir (the dot-dashed line), Tenerife (the dotted line) \& Namibia (the dot-dot-dashed line). (a) Trigger rates without the GF; the rates differ only due to the difference of altitudes. (b) Trigger rates for the Northern sites with the minimum and maximum GF at $\theta=30^{\circ}$; in upper (thicker) lines are for $\phi = 0^{\circ}$ and the lower (thinner) lines are for $\phi = 180^{\circ}$. (c) Trigger rates for the Southern sites with the minimum and maximum GF at $\theta=30^{\circ}$; the upper (thicker) lines are for $\phi = 180^{\circ}$ and the lower (thinner) lines are for $\phi = 0^{\circ}$.}
\label{fig:rates}
\end{center}
\end{figure}

We analyse the images obtained from {\fontfamily{pcr}\selectfont Sim\_telarray} simulations using the {\fontfamily{pcr}\selectfont read\_cta} program (which is an internal component of the CTA simulation package). After the image cleaning, for which we use the tail-cuts\footnote{Pixels are considered to be a part of an image and are called boundary/core pixels if their signal is larger than 5.5/11 pe and have at least one neighbouring pixel with charge above 11 pe.} of 5.5/11 photo-electrons (pe) (for cleaning algorithm details, see Sec.~4 of \citep{Daum97}), we get the image parameters discussed in Sec.\  \ref{sec:images}.

\section{{\bf Results}}
\label{sec:results}

\subsection{Trigger rates, collection areas and energy thresholds}
\label{sec:4.1}
In this section we present several {\it trigger level} parameters, which can be used to describe the performance of ground-based gamma ray detectors, cf.\ \citep{hinton09}.

\begin{figure}[t]
\begin{center}
\includegraphics[trim = 8mm 0mm 0mm 1mm, clip,scale=0.44]{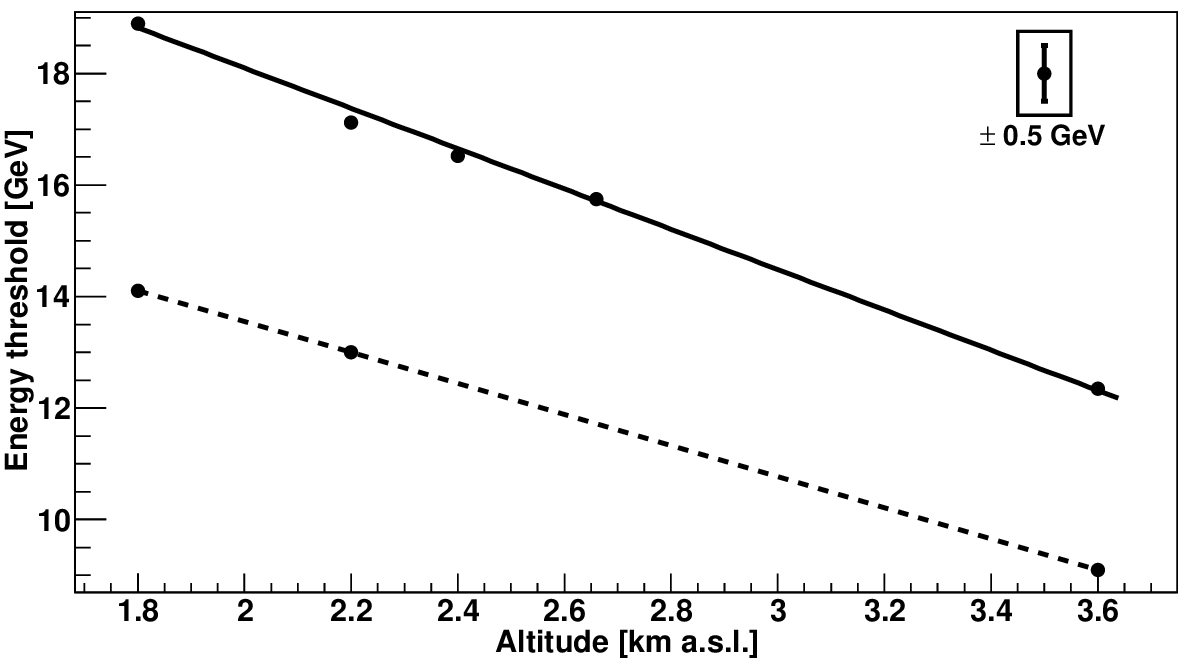}
\caption{Energy threshold at the sites with neglected GF; the thresholds differ only due to the difference of altitudes between the sites. The upper set of points is for $\Gamma=2.6$ and the points (from right to left) correspond to Argentina-Salta, Argentina-Leoncito, San Pedro Martir, Tenerife and Namibia. The lower set of points is for $\Gamma=3.5$ and the points are (from right to left) for Argentina-Salta, Tenerife and Namibia. The vertical bar in the frame represents the errors of energy threshold values ($\pm 0.5\; {\rm GeV}$).}
\label{fig:et_h}
\end{center}
\end{figure}

\begin{figure}[t!]
\begin{center}
\includegraphics[trim = 8mm 3mm 0mm 1mm, clip, scale=0.45]{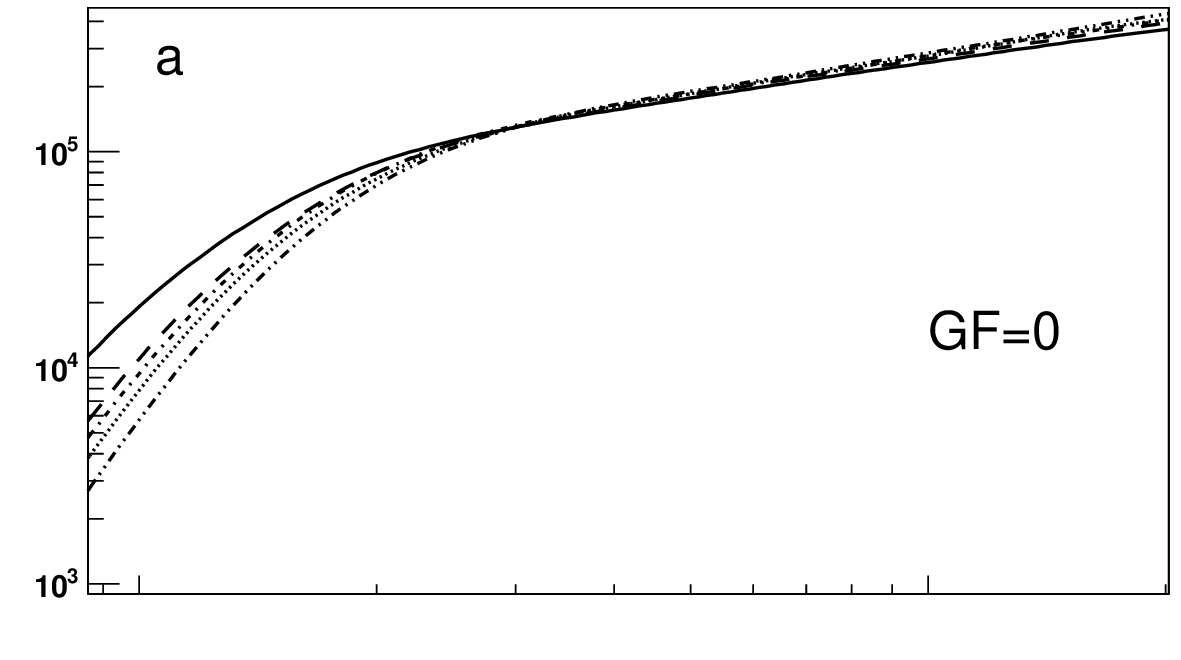}
\includegraphics[trim = 8mm 3mm 0mm 1mm, clip, scale=0.45]{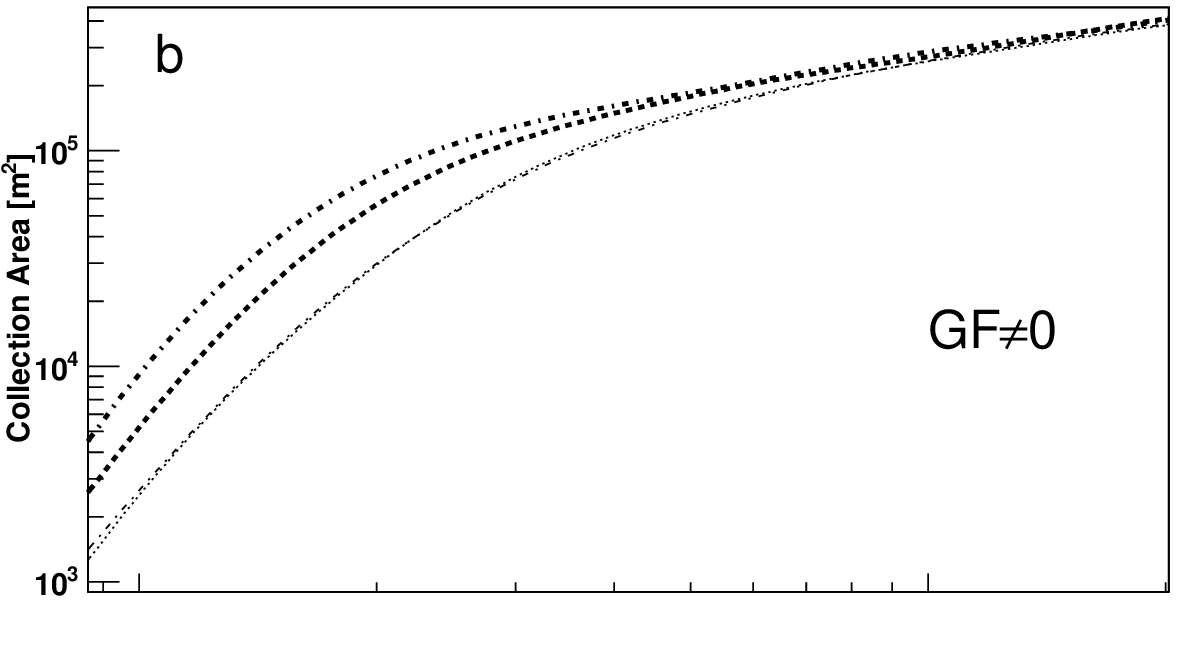}
\includegraphics[trim = 8mm 3mm 0mm 1mm, clip, scale=0.45]{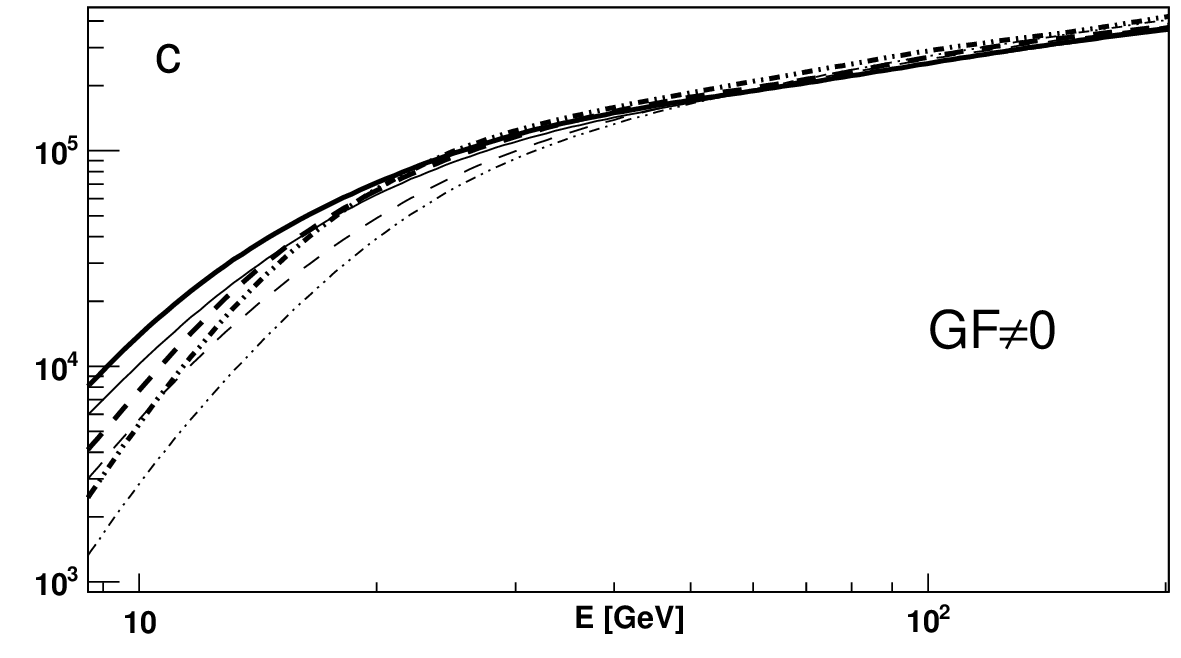}
\caption{Collection area (smoothed histograms) as a function of energy for Argentina-Salta (the solid line), Argentina-Leoncito (the dashed line), San Pedro Martir (the dot-dashed line), Tenerife (the dotted line) \& Namibia (the dot-dot-dashed line). (a) Collection areas at all 5 sites without the GF. (b) Collection areas for the Northern sites with the minimum and maximum GF at $\theta=30^{\circ}$; in upper (thicker) lines are for $\phi = 0^{\circ}$ and the lower (thinner) lines are for $\phi = 180^{\circ}$. (c) Collection areas for the Southern sites with the minimum and maximum GF at $\theta=30^{\circ}$; the upper (thicker) lines are for $\phi = 180^{\circ}$ and the lower (thinner) lines are for $\phi = 0^{\circ}$.}
\label{fig:areas}
\end{center}
\end{figure}

The probability of a detection of gamma rays with energy $E$ is given by the effective collection area, $A(E)$, determined in MC simulations  by the range of the impact parameter, $D$, and the ratio of triggered, $n_{\rm tr}$, and simulated, $n_{\rm sim}$, events, $A(E) = [n_{\rm tr}(E)/n_{\rm sim}(E)] \pi D^2$.
We calculate also the differential trigger rates, $R(E)$, related to $A(E)$ by $R(E) = \Phi(E)\times A(E)$, where $\Phi(E)$ is the primary gamma ray differential energy spectrum. To set the normalization of $R(E)$, we assume the primary differential energy spectrum of $\Phi(E)=2.8 \times 10^{-13} (E/1 {\rm TeV})^{-2.6}$ cm$^{-2}$ s$^{-1}$ TeV$^{-1}$, i.e., at 1 TeV, 1\% of the Crab Nebula flux (cf.\ \citep{Aharonian01}); we discuss in detail our results for such a primary spectrum and then we briefly compare them with the results for a softer spectrum, with $\Gamma=3.5$. Following the conventional definition for IACT detectors, we determine the  energy threshold, $E_{\rm th}$, as the peak energy of the differential trigger rate\footnote{Technically, the energy peak is determined by the mean of a Gaussian fit.}.

To illustrate the altitude effect (cf.\ \citep{Aharonian01,Konopelko04}), we show in Fig.\ \ref{fig:rates}a the differential trigger rates for the sites with neglected GF. Below 30 GeV,  the trigger rates increase with increasing   altitude due to two effects. First, the atmospheric transmission for Cherenkov 
light is higher,  as the air mass between the shower maximum and the telescope is lower, at higher altitudes. Second, the light pool area is smaller (an obvious geometrical effect) and, hence, the Cherenkov photons density at distances close to the core axis is higher at higher altitudes. These two effects are important only for low energy showers, which produce  Cherenkov light intensities not exceeding significantly the threshold level for triggering a telescope. We can see in Fig.\ \ref{fig:rates}a that the peak energy decreases with increasing altitude. For the telescope
parameters and sites assumed in this work, the dependence of the
energy threshold on the altitude of the array can be well described by
a linear relation, see Fig.\ \ref{fig:et_h}, with 
\begin{align}
E_{\rm th}^0(h) = [25.4 - 3.6 \times (h / 1 {\rm km})]\; {\rm GeV}, 
\label{eq:E_h}
\end{align}
where $h$ is the altitude; superscript '0' denotes the zero magnetic field.
Extrapolating it to higher altitudes we get $E_{\rm th}^0 = 7$ GeV at 5 km a.s.l., in approximate agreement with \citep{Aharonian01}. The dependence of the collection area on altitude (see Fig.~\ref{fig:areas}a) is significant only around the threshold and at sub-threshold energies, e.g.\ at 16 GeV the collection areas at 1.8 km and 3.6 km differ by a factor of 3, but already at 20 GeV the difference decreases to 30\%. 

Above 30 GeV, without GF, the differential trigger rates decrease, and the corresponding collection  areas are smaller, with the increase of altitude. Primary photons with such energies produce a sufficiently high  Cherenkov photon density to trigger a telescope and their detection is not much affected by atmospheric absorption or changes of photon density with altitude. The Cherenkov light pool is smaller at higher altitudes and, as a consequence, the detection probability for high energy gamma rays is lower.
This is, however, a minor effect, e.g.\ at 1 TeV the collection areas for the largest (Salta) and smallest (Namibia) altitudes differ by only 20\%. Moreover, this effect will not necessarily affect the full CTA array, with small size telescopes covering a much larger area; such a more extended array should be less sensitive to changes of the light-pool size with altitude.

Overall, our results are consistent with previous studies of the influence of altitude on the CTA performance, see fig.~19 in \citep{cta}. 

To complete our discussion of the altitude effect, we briefly consider the array spacing which, after optimization for the height of a site, may improve the performance parameters of the IACT \citep{Aharonian01,Konopelko04}. Assuming that the LSTs separation of $105$ m, considered in this work, is optimal for sites at the altitude of $2000$ m a.s.l.~(see Chapt.~8 in \citep{cta} or Chapt.~6 in \citep{Bernlohr12}), and that the light-pool radius scales linearly with the distance to the shower maximum, the optimal spacing of $82$ m can be estimated for Salta (which is the highest-altitude site considered here). We performed simulations  for array E(L) with such a spacing at the Salta site and we found that the total trigger rate is larger by $\approx 2\%$, the energy threshold is reduced by less than 1 GeV (i.e.\ the difference is comparable to the precision, $\pm 0.5\; {\rm GeV}$, of determining the value of $E_{\rm th}$) and  the collection area increases by less than 20\% (and only for $E < 30\; {\rm GeV}$) with respect to our results for the 105 m spacing at this site. This confirms that a slight improvement of the low-energy performance of IACT can be achieved by an optimization of the spacing and the detailed study focused on this issue should be performed after the site-selection decisions are made. However, the spacing effect is weaker than those studied in this paper and our simplified assumption of the same spacing at each site does not affect our overall conclusions.

Similarly to the altitude, the GF effect reduces significantly the trigger rates and the collection areas, see Figs.\ \ref{fig:rates}b,c and \ref{fig:areas}b,c, only in the low energy range; for higher energy of primary photons, the Cherenkov light density is sufficient even if the Cherenkov pool is distorted by the GF effect.  For $B_{\perp} \simeq 40 \;{\rm \mu T}$, corresponding to $\phi = 180^{\circ}$ in the northern sites,
lower curves in Figs.~\ref{fig:rates}b and \ref{fig:areas}b, it has a noticeable effect up to $\approx 100$ GeV; for these values of $B_{\perp}$, the collection areas at 20 GeV are reduced by almost a factor of 3 with respect to the corresponding values without the GF. 
For the Southern sites, changes in $R(E)$ and $A(E)$ resulting from the GF effect are  significant up to $E \approx 60$ GeV. 

\begin{figure}[t]
\begin{center}
\includegraphics[trim = 8mm 4mm 0mm 1mm, clip,scale=0.44]{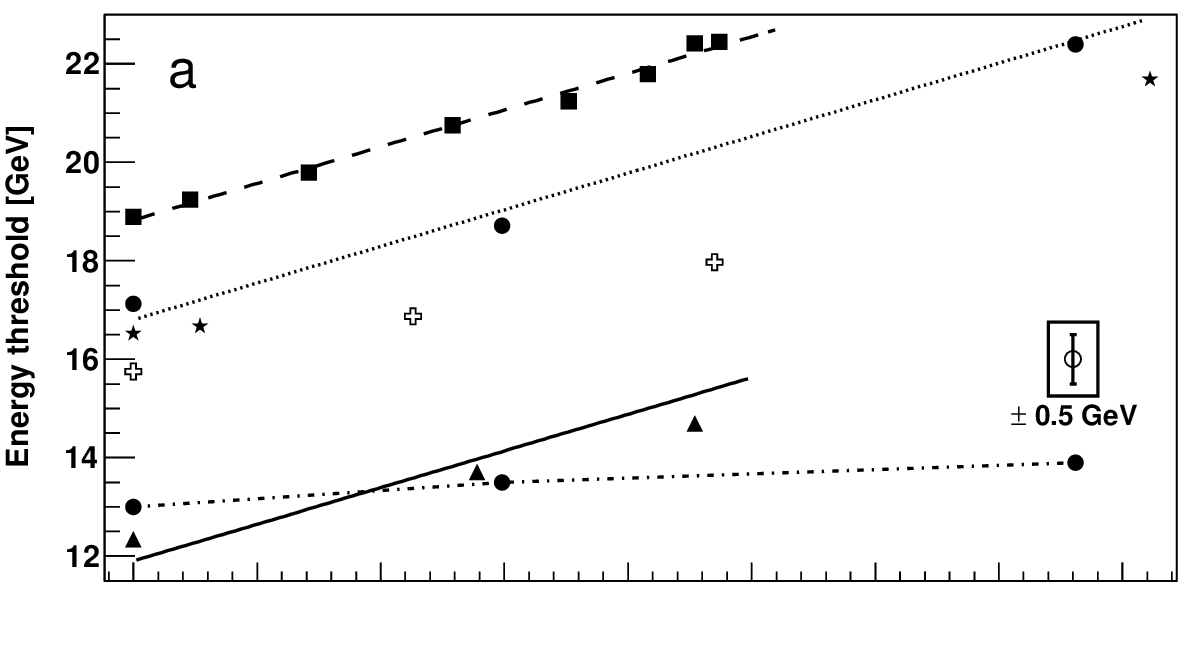}
\includegraphics[trim = 8mm 2mm 0mm 1mm, clip,scale=0.44]{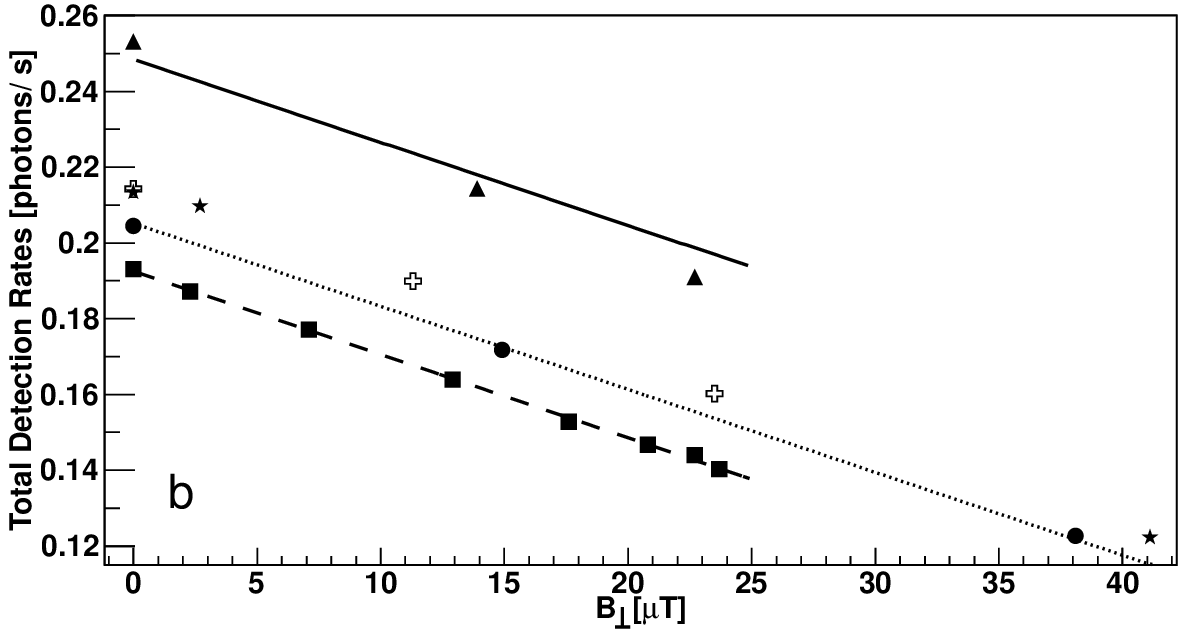}
\caption{Energy threshold (top panel) and the total gamma trigger rates for a pure power-law, ${E}^{-2.6}$ spectrum with flux at 1 TeV equal to 1\% of the Crab source (bottom panel), both as a function of $B_{\perp}$ for Argentina-Salta (triangles \& solid line), Argentina-Leoncito (crosses), San Pedro Martir (stars), Tenerife (circles) and Namibia (boxes). The lines with the slope from the linear fit to the Namibian site are shown for selected sites. The lower set of circles in panel (a), connected by the dot-dashed line, are for Tenerife and $\Gamma=3.5$, all the remaining points correspond to $\Gamma=2.6$. The vertical bar in the frame, in the top panel represents the errors of energy threshold values ($\pm 0.5\; {\rm GeV}$). The statistical error of the total detection rates $(\approx 0.001 {\rm phot/s})$ is approximately half a size of the markers in the bottom panel.}
\label{fig:et_tr_b}
\end{center}
\end{figure}

Fig.\ \ref{fig:et_tr_b}a summarizes our results on the energy threshold for all cases calculated in our work. Fig.\ \ref{fig:et_tr_b}b shows a similar summary for the total trigger rate, $R_{\rm tot}$, obtained by integrating $R(E)$ over the total energy range range (from 3 GeV to 1 TeV). Both quantities are shown as functions of $B_{\perp}$ 
(including the cases of null $B$), rather than the azimuthal angles at the sites, as we aim to derive more general properties. Remarkably, both quantities appear to scale linearly with ${B}_{\perp}$ at a fixed altitude.  By fitting a linear function to the data points for the Namibian site we get the slopes of these relations; the altitude-dependent constant follow from the results with neglected GF.

For the threshold energy we get
\begin{align}
E_{\rm th} = [0.15 \times (B_{\perp} / 1 {\rm \mu T}) + E_{\rm th}^0(h) ]\; {\rm GeV},
\label{eq:E_th}
\end{align}
where $E_{\rm th}^0$ is given by Eq.\ (\ref{eq:E_h}).
The fitted slope gives the correct extrapolation up to 40 ${\rm \mu T}$, see the dotted line in Fig.\ \ref{fig:et_tr_b}a. On the other hand, we note some hints for flattening of this relation for larger $h$, e.g.\ the three data points for Salta indicate the slope of 0.1 GeV/${\rm \mu T}$ (even it is lower for Leoncito); however, deviations from the relation determined for Namibia do not exceed the precision, $\pm 0.5$ GeV, of determining the peak position in $R(E)$.

For the trigger rate we get
\begin{align}
R_{\rm tot} = [R_{\rm tot}^0(h) - 0.0022 \times (B_{\perp} / 1 {\rm \mu T})]\;{\rm phot/s}, 
\label{eq:R_tot}
\end{align}
with 
\begin{align}
R_{\rm tot}^0(h) = [0.033 \times (h / 1 {\rm km}) + 0.13]\;{\rm phot/s}. 
\label{eq:R_tot2}
\end{align}
Again, we find that this relation works well up to at least 40 ${\rm \mu T}$. We also find that the same slope describes correctly all sites except for Salta, for which we find a slight steepening with the slope of 0.0027 phot/s/${\rm \mu T}$.

All the above results correspond to $\Gamma=2.6$. For $\Gamma=3.5$ we get a significantly lower $E_{\rm th}$, by 4 up to 9 GeV (depending on $h$ and $B_{\perp}$), furthermore, the slopes of the linear functions describing the $E_{\rm th}(h)$ and $E_{\rm th}(B_{\perp})$ dependence are different than for $\Gamma=2.6$ (see Fig.~\ref{fig:et_h} and \ref{fig:et_tr_b}). Specifically, the dependence on $B_{\perp}$ is reduced for a softer primary spectrum and for $\Gamma=3.5$ the $E_{\rm th}(B_{\perp})$ function is almost flat, as shown by the dot-dashed line in Fig.~\ref{fig:et_tr_b}a. The dependence on $h$ is also weaker than for $\Gamma=2.6$ and the difference between $E_{\rm th}$ for various $\Gamma$ decreases with increasing altitude (Fig.~\ref{fig:et_h}). On the other hand, the total trigger rates are similar for both $\Gamma=2.6$ and 3.5, if we assume that both primary spectra have the same normalization around the energy threshold (i.e.~at $\simeq 20$ GeV) and, therefore, $R_{\rm tot}$ for $\Gamma=3.5$ are not shown in Fig.~\ref{fig:et_tr_b}b for clarity of the figure.

The integrated trigger rates are determined much more precisely than the peak position of the differential rates. Taking $\sqrt{n_{\rm tr}}$ as a measure of the statistical error, we get the relative error of only $\Delta R_{\rm tot}/R_{\rm tot} \approx 4\times10^{-3}$. Therefore, we regard the slopes of our linear fits determined by three data points for sites other than Namibia to be more reliable for the $R_{\rm tot}(B_{\perp})$ relations than for $E_{\rm th}(B_{\perp})$.
Deviations from the slopes of our fits for the Namibian site indicate an opposite effect of large magnetic field at large altitudes, namely, a stronger and weaker reduction of the IACT performance, for $R_{\rm tot}$ and $E_{\rm th}$, respectively.
Again, we regard the former property to be more reliable.

We note also that the values of $R_{\rm tot}$ and $E_{\rm th}$ for array K(L) (with 5 telescopes) are systematically different from those presented in Fig.\ \ref{fig:et_tr_b} for array E(L). Specifically, $R_{\rm tot}$ is larger by a factor of 1.2, however, the slopes of $R_{\rm tot}(B_{\perp})$ and $E_{\rm th}(B_{\perp})$ are the same as for array E(L).

\subsection{Image parameters}
\label{sec:images}

The separation of the gamma ray signal from the dominating hadronic background is a fundamental issue in the IACT technique. 
An effective way to separate gamma rays from the background exploits the differences in the distributions
of the image parameters for gamma and hadronic showers. Those differences are used by most of gamma/hadron separation techniques, like scaled cuts and Random Forest (e.g., \citep{albert08b,konopelko99}).
In this section we consider the GF and altitude effects on the two, most sensitive, separation parameters characterising the shape and orientation of shower images.

The Hillas width and length parameters \citep{Hillas85}, defined as the second central moments calculated
along the minor and major axes of the image, are good separation parameters, at least for large image sizes\footnote{the size is defined as the total integrated light of the shower image}. Namely, the mean width and length as functions of size (referred to as the width and length profile) are different for gamma rays and hadrons.
\begin{figure}[t]
\begin{center}
 \scalebox{0.95}{%
  \includegraphics[width=4.6cm,height=4.6cm]{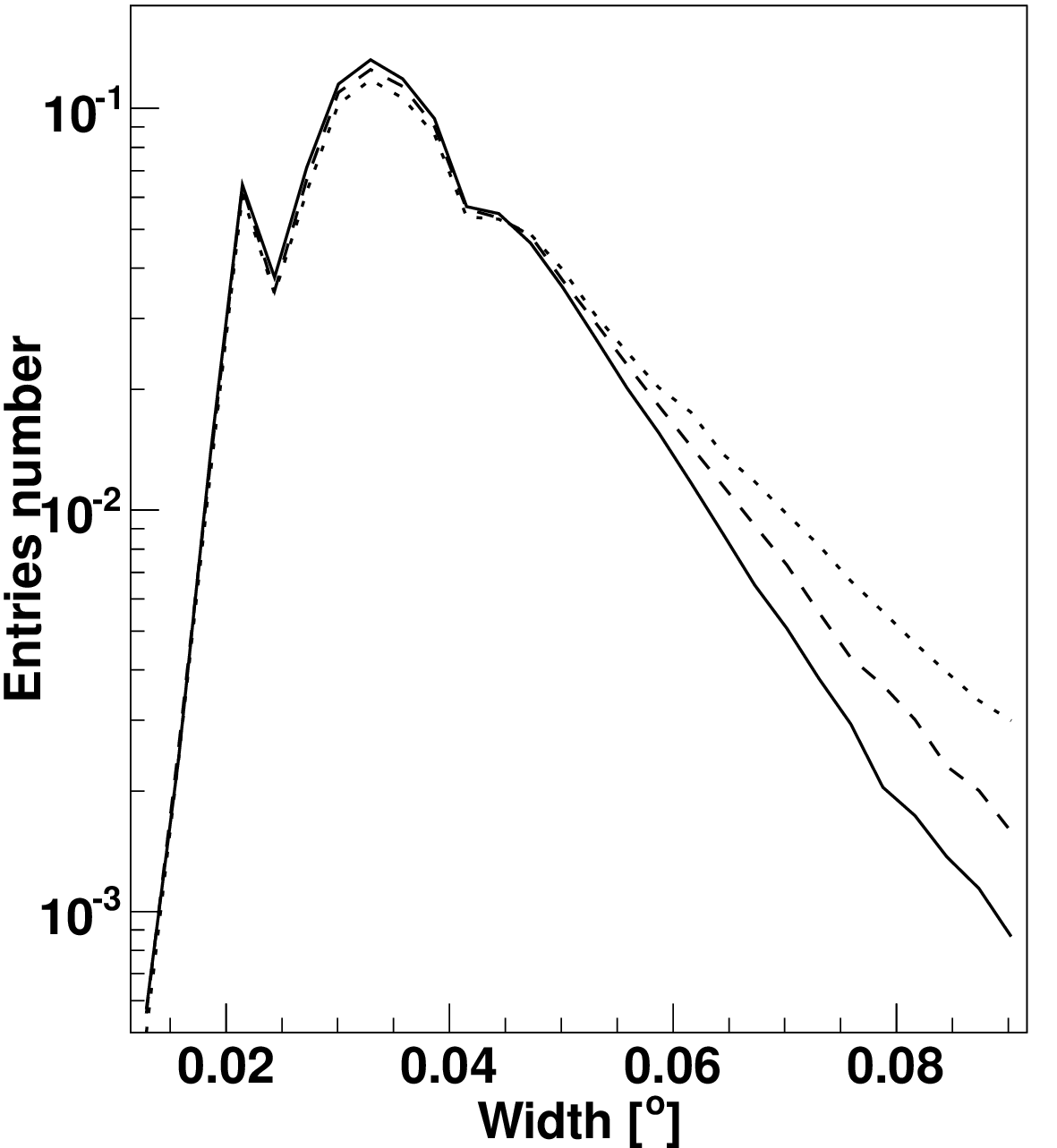}}%
 \scalebox{0.95}{%
  \includegraphics[width=4.6cm,height=4.6cm]{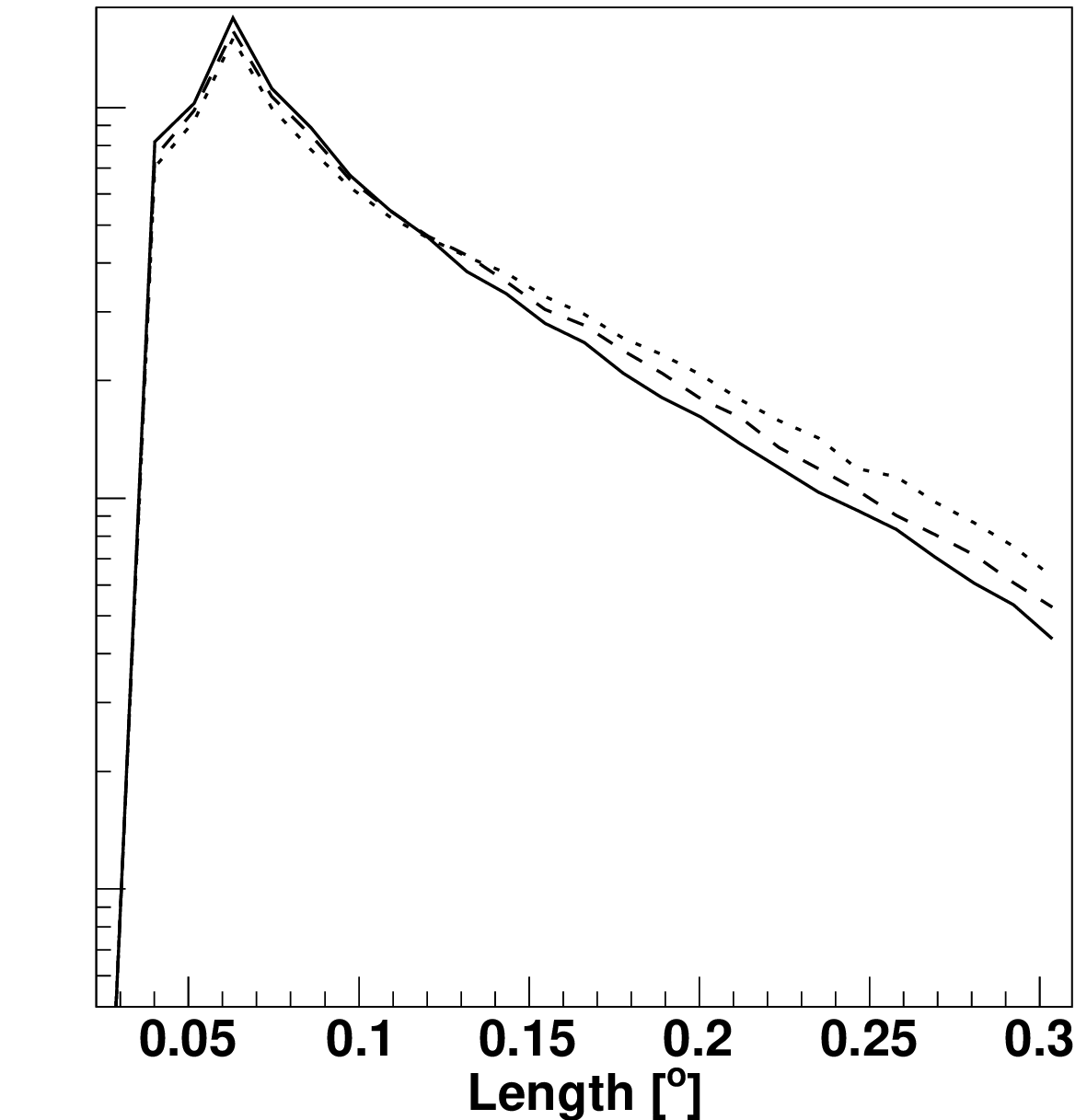}}%
\caption{The distributions of the Hillas width (left) and length (right) parameters from our gamma ray simulations for the Tenerife site.
 In both figures the curves from bottom to top correspond to $B=0$ (solid), $\phi=0^{{\circ}}$ (dashed) and $\phi=180^{{\circ}}$ (dotted).
Multiple peaks in the width distributions appear due to the pixelization of the camera. The peaks' sharpness is a result of histograms binning.}
\label{fig:wl}
\end{center}
\end{figure}

The broadening of images by the GF effect should be reflected in the length and width distributions. 
Indeed, as can be seen in Fig.~\ref{fig:wl}, 
the relative number of gamma ray images with large lengths and widths increases with increasing $B_{\perp}$; the effect seems to be stronger for the latter, then, we focus below on the width profiles.
At the width values $< 0.05^{\circ}$, their distribution appears to be not affected by the GF, which is an obvious (camera-dependent) effect for the camera pixel size of $0.09^{\circ}$.

\begin{figure}[t]
\begin{center}
\includegraphics[trim = 8mm 3mm 0mm 2mm, clip,scale=0.44]{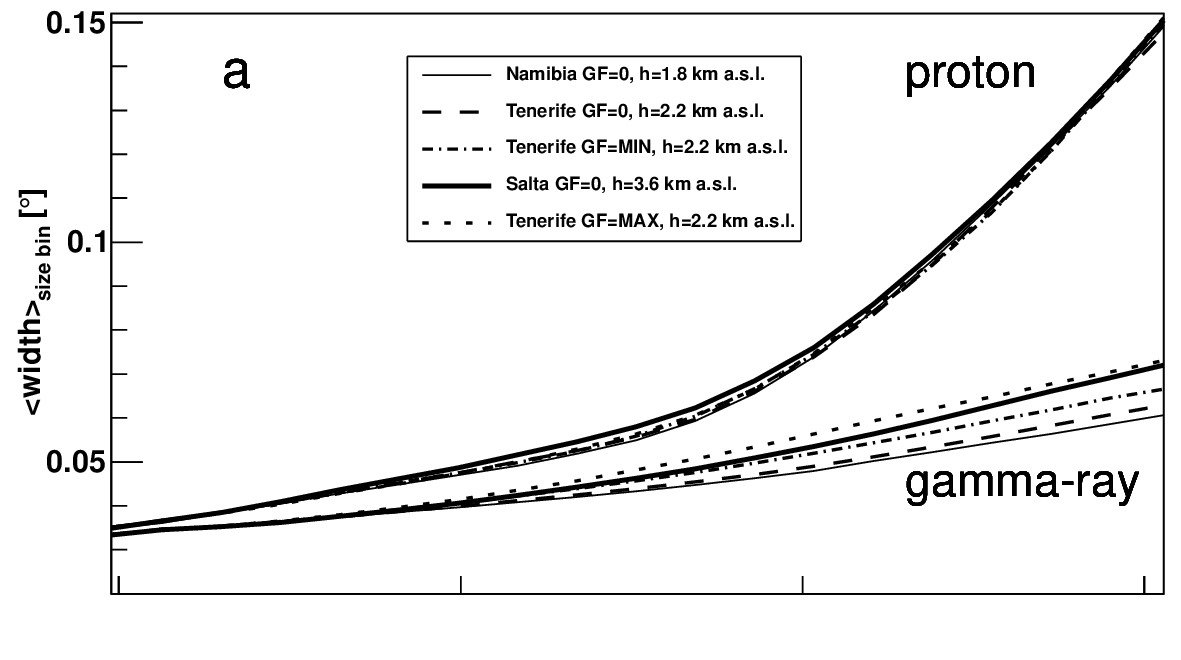}
\includegraphics[trim = 8mm 3mm 0mm 1mm, clip,scale=0.44]{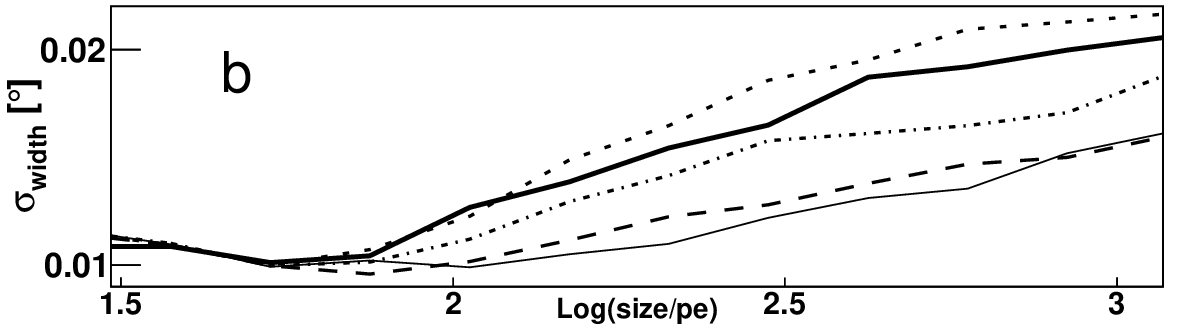}
\caption{The mean width for gamma rays and protons (a) and standard deviation for gamma rays (b) as the function of size for Namibia without GF (thinner solid), Tenerife without GF (dashed),  Tenerife with GF at $\phi = 0^{\circ}$ (dot-dashed), Argentina-Salta without GF (thicker solid) and Tenerife with GF at  $\phi = 180^{\circ}$ (dotted) from bottom to top in both panels. The profiles for Argentina and Namibia illustrate the effect of changing the altitude only; the gamma ray profiles for Tenerife illustrate the effect of changing GF at a fixed altitude.}
\label{fig:ws}
\end{center}
\end{figure}

Fig.~\ref{fig:ws} shows the width profiles obtained from our simulated data 
by dividing the size range into bins and calculating the mean width, $<$width$>$, and standard deviation, $\sigma_{\rm width}$, of the width distribution for each bin. For gamma rays, the increase of $B_{\perp}$ at a fixed $h$ leads to the increase of $<$width$>$ at sizes exceeding 100 pe,  e.g.\ $<$width$>$ increases by $\sim 20\%$ with the increase of $B_{\perp}$ from 0 to $40\; {\rm \mu T}$. The increase of $<$width$>$ is accompanied by a similar in magnitude increase of standard deviation ($\sigma$), see Fig.~\ref{fig:ws}b. 
Neglecting the GF effect, we find that for gamma rays the mean width, for a given size bin, increases with increasing altitude, see Fig.~\ref{fig:ws}(a), which effect results simply from the decreasing distance between the telescope and the shower maximum, see \citep{Konopelko04}.
Note that the magnitudes of the changes of the gamma ray width profiles due to both effects (i.e.\ GF and altitude) are similar. 
The changes are not dramatic, however, neglecting them may result in systematic inaccuracies e.g.\ in the scaled cuts technique, discussed in Section \ref{sec:G/H_sep}.

Another elementary information, which must be taken into account for gamma/hadron discrimination, concerns the  shower direction. In the stereo (or higher multiplicity) IACT systems, the arrival direction of the primary particle is estimated as the crossing point of the main axes of images collected from the individual telescopes, see \citep{Fegan97}.
In our analysis, the reconstructed shower direction is determined by the weighted average of all intersection points (i.e.\ for all pairs of triggered telescopes), according to 
the  scheme described in \citep{Bernlohr12}. 
Then, we determine the $\mathit{\Theta}$ image parameter, defined as the angular distance between the simulated and the reconstructed directions of the primary gamma ray.

As pointed out in previous studies, the reconstruction accuracy is affected by both the GF strength \citep{Commichau08}  and the altitude \citep{Konopelko04}. The former effect involves 
  the change of the orientation of the individual shower images by the GF, which obviously spoils the direction reconstruction. The altitude effect, studied in detail in \citep{Konopelko04,Sobczynska09}, involves the radial distribution of the photon density in the light pool.
Namely, the increase of the Cherenkov light density at the higher-altitude sites (underlying the enhanced trigger efficiency, studied in the previous section) occurs mostly close to the shower axis and, therefore, showers with smaller impact parameters are more frequently recorded. However, images of such showers have more circular shape and the orientation of their major axes tends to be poorly determined.

\begin{figure}[t]
\begin{center}
 \scalebox{0.95}{%
  \includegraphics[width=4.6cm,height=4.6cm]{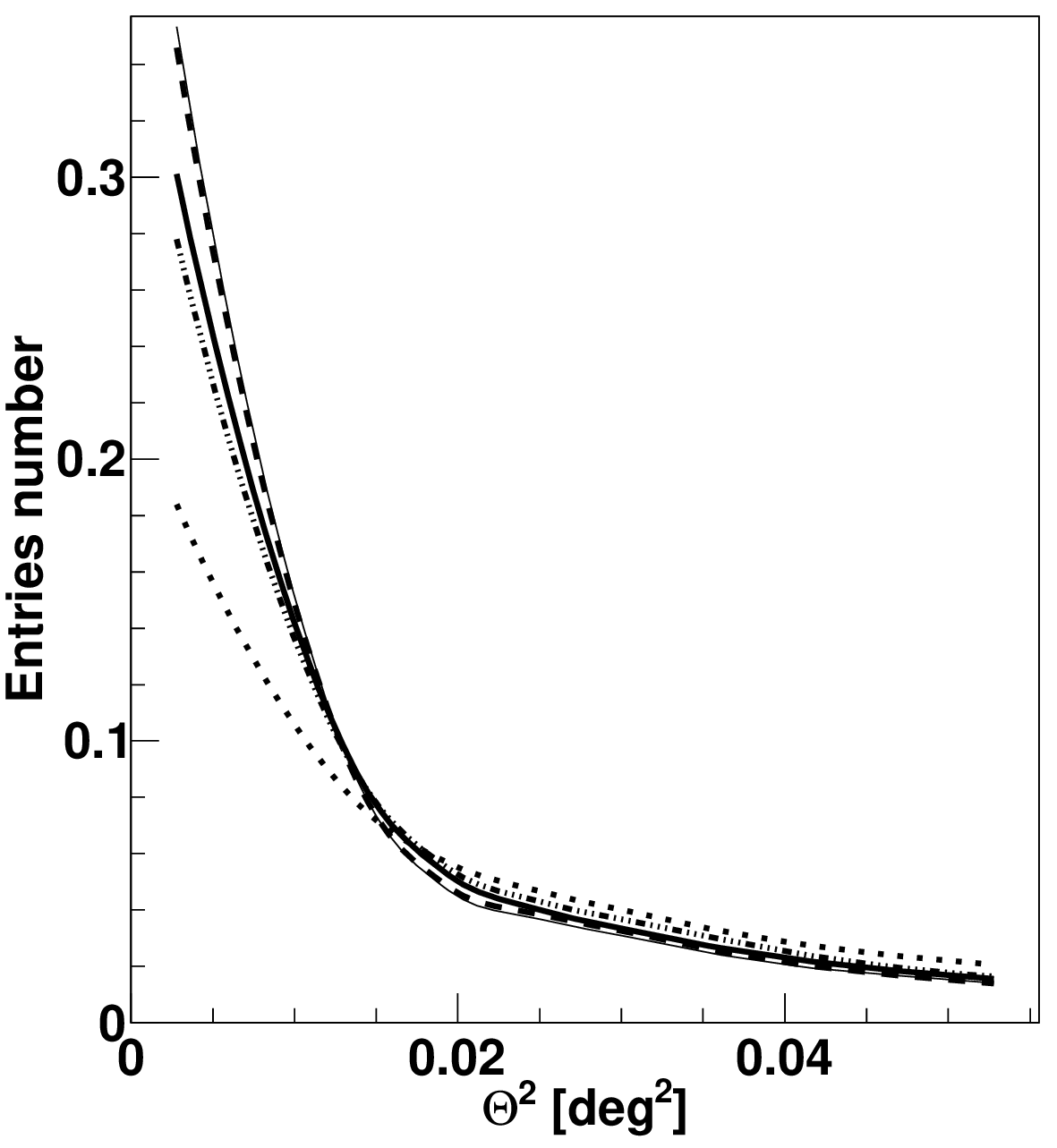}}%
 \scalebox{0.95}{%
  \includegraphics[width=4.6cm,height=4.6cm]{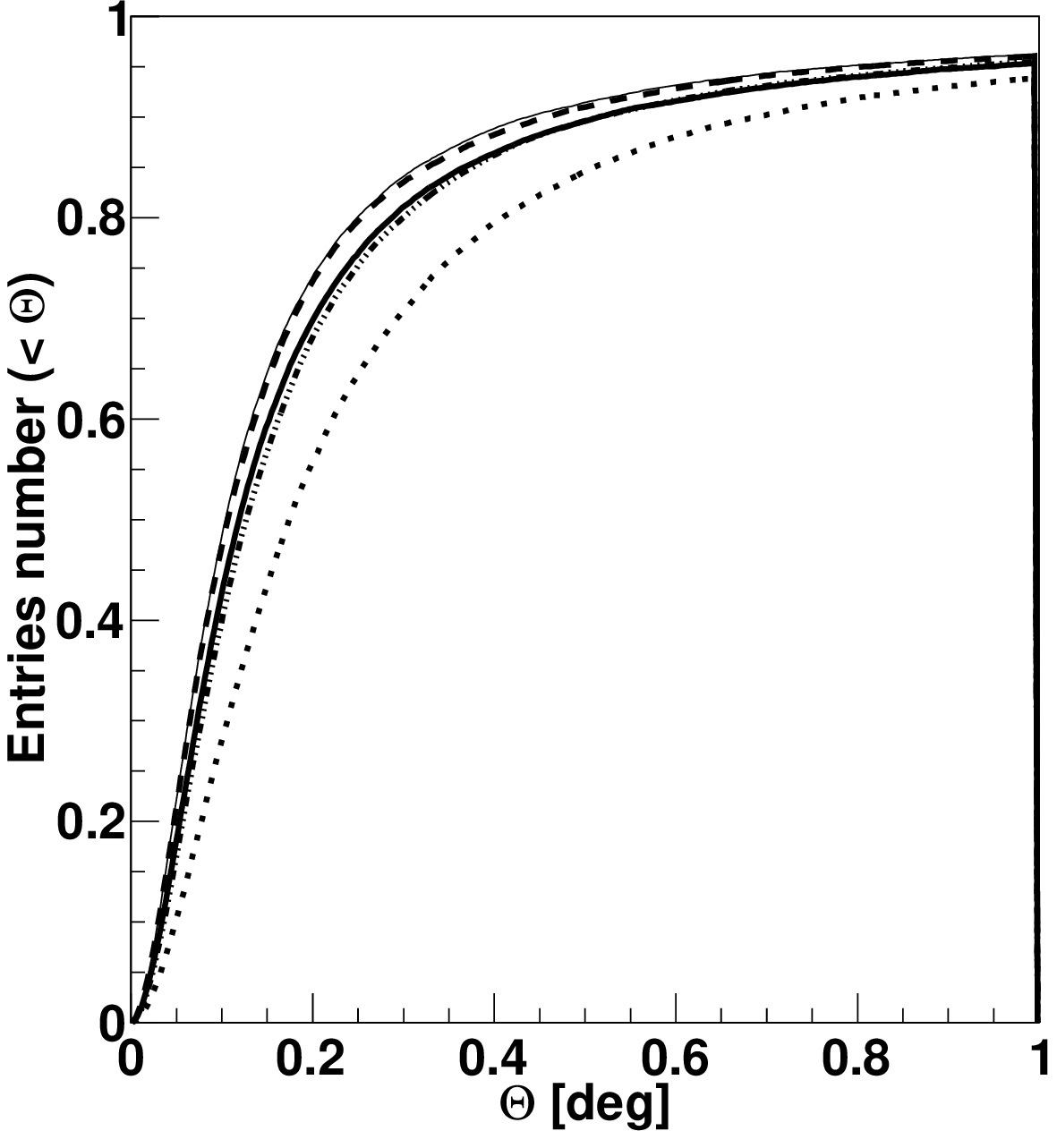}}%
\caption{$\mathit{\Theta}^2$ distribution for our MC gamma simulations  (left) and the corresponding cumulative $\mathit{\Theta}$ (right)  for  Namibia without GF (thinner solid), Tenerife without GF (dashed), Argentina-Salta without GF (thicker solid), Tenerife with GF at $\phi = 0^{\circ}$ (triple dot-dashed) and Tenerife with GF at  $\phi = 180^{\circ}$ (dotted). Only events with $Size > 100\,$ pe are used for both plots.}
\label{fig:rec}
\end{center}
\end{figure}

The left panel in  Fig.~\ref{fig:rec} shows the  $\mathit{\Theta}^2$ plots (the distribution of the squared distance between the simulated and the reconstructed source position) for  images with {\it Size} $>100$ pe; images with smaller sizes tend to have poorly determined orientation and are less suitable to illustrate effects in the reconstruction quality.
In agreement with \citep{Commichau08} and \citep{Konopelko04}, our  $\mathit{\Theta}^2$ plots  indicate that the most accurate direction reconstruction is achieved for the lowest altitude without GF. We can see also that the altitude effect in the $\mathit{\Theta}^2$ plots is much weaker than the GF effect.  E.g.~the number of events reconstructed within $\mathit{\Theta}^{2}<0.03\; {\rm deg^{2}}$ is reduced by $\approx30\%$ due to the change of $B_{\perp}$ from 0 to $40 \;{\rm \mu T}$, and only by 10\% due to the increase of altitude from 1.8 to 3.6 km a.s.l.. This can be easily understood as the altitude affects the accuracy of determining the image orientation but the actual orientation is not changed, whereas the GF does change the orientation.

Obviously, the precision of the direction reconstruction determines the angular resolution, briefly discussed in Section \ref{sec:angular_res}.

\subsection{The hadronic background}
\label{sec:prot}

To check how the geophysical conditions affect the hadronic-events detection, we applied the analysis steps described in the previous sections to our MC simulations of hadronic showers for Salta, Namibia and Tenerife.
At the trigger level, the detection efficiency of protons is  much less sensitive to changes of both the GF and the altitude than the detection of gamma rays. Without the GF, the total trigger rate of protons at Salta is only by 2\% smaller than at the HESS-site. For Tenerife, the increase of $B_{\perp}$ from 0 to $40\; {\rm \mu T}$ results in the decrease of the trigger rate of protons by 10\% (as compared to the 40\% decrease for gamma rays).

Also the protonic images, being intrinsically broader than the gamma images, are much less affected by both the GF and the altitude. In particular, their influence on the protonic width profiles appears insignificant, see Fig.~\ref{fig:ws}a.

\subsection{Gamma/hadron separation}
\label{sec:G/H_sep}

In this section we consider the influence of the GF and altitude on the two most efficient discrimination parameters, i.e.\ the scaled width (defined below) and $\mathit{\Theta}^2$. We investigate the separation quality at Salta, Namibia
and Tenerife, as measured by the quality factor 
\begin{align}
QF\equiv \frac{ n^{\gamma}_{\rm cuts}/n^{\gamma}_{\rm tr}}{\sqrt{n^{h}_{\rm cuts}/n^{h}_{\rm tr}}},
\label{eq:qf}
\end{align}
where $n^{\gamma,h}_{\rm tr}$ and $n^{\gamma,h}_{\rm cuts}$ are the number of gamma (hadron) events that passed the image cleaning and a given separation procedure, respectively.

The scaled cuts technique uses the widths (or, similarly, lengths) scaled to values expected for gamma rays. In a given size range, the scaled width distribution is computed as $w_{\rm s}\equiv (width\;-<\!w\!>)/\sigma_{\rm width}$, where $<\!w\!>$ and $\sigma_{\rm width}$ are the mean width and standard deviation, as defined in Section \ref{sec:images}. Given the difference in the gamma and hadron {\it width} distributions illustrated in Fig.~\ref{fig:ws}, even a simple cut, e.g.\ accepting only events with $w_{\rm s}<1$, results in efficient rejection of hadrons (except  for small sizes).

\begin{figure}[t]
\begin{center}
 \scalebox{0.95}{%
  \includegraphics[width=4.6cm,height=4.6cm]{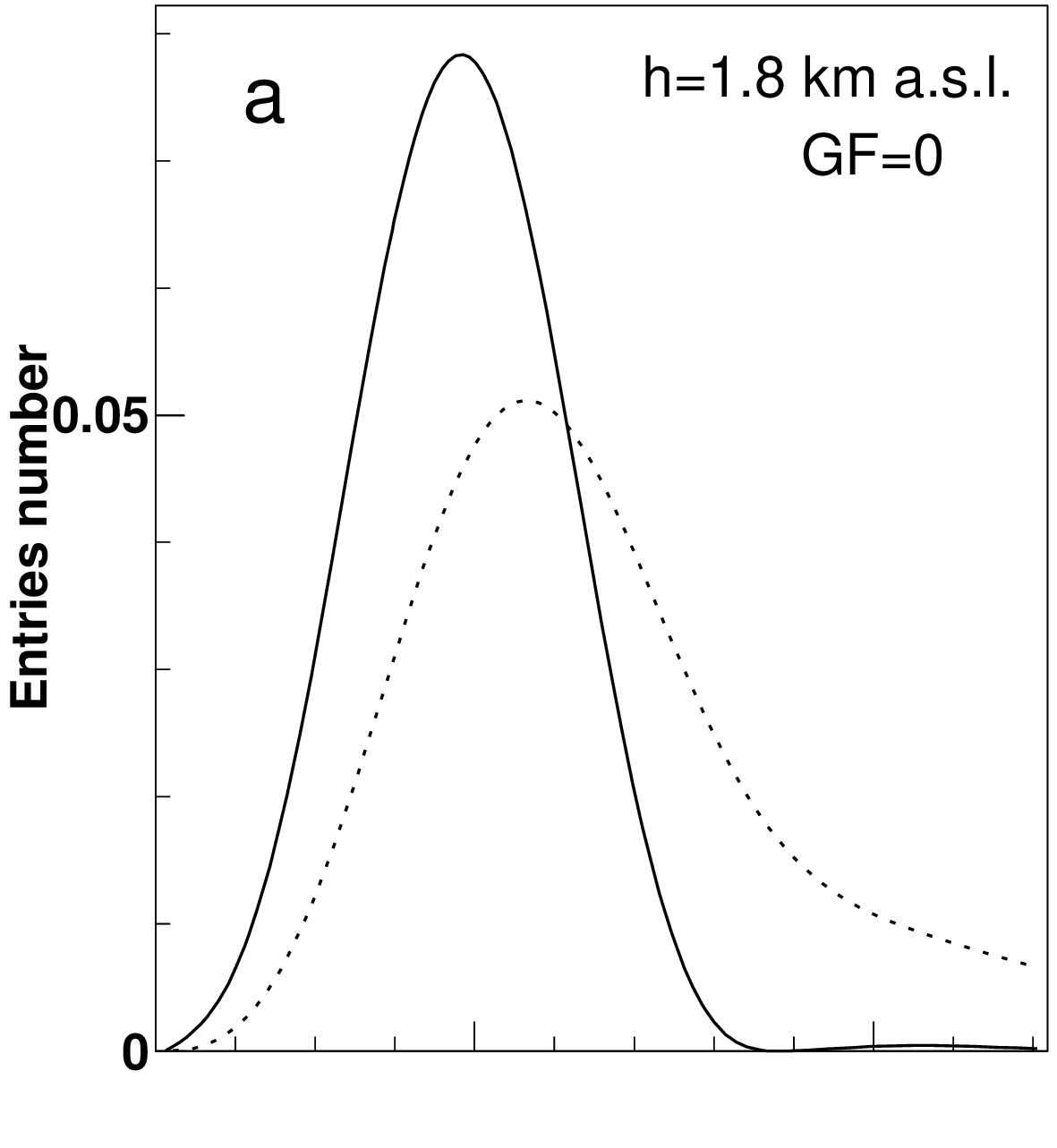}}%
 \scalebox{0.95}{%
  \includegraphics[width=4.6cm,height=4.6cm]{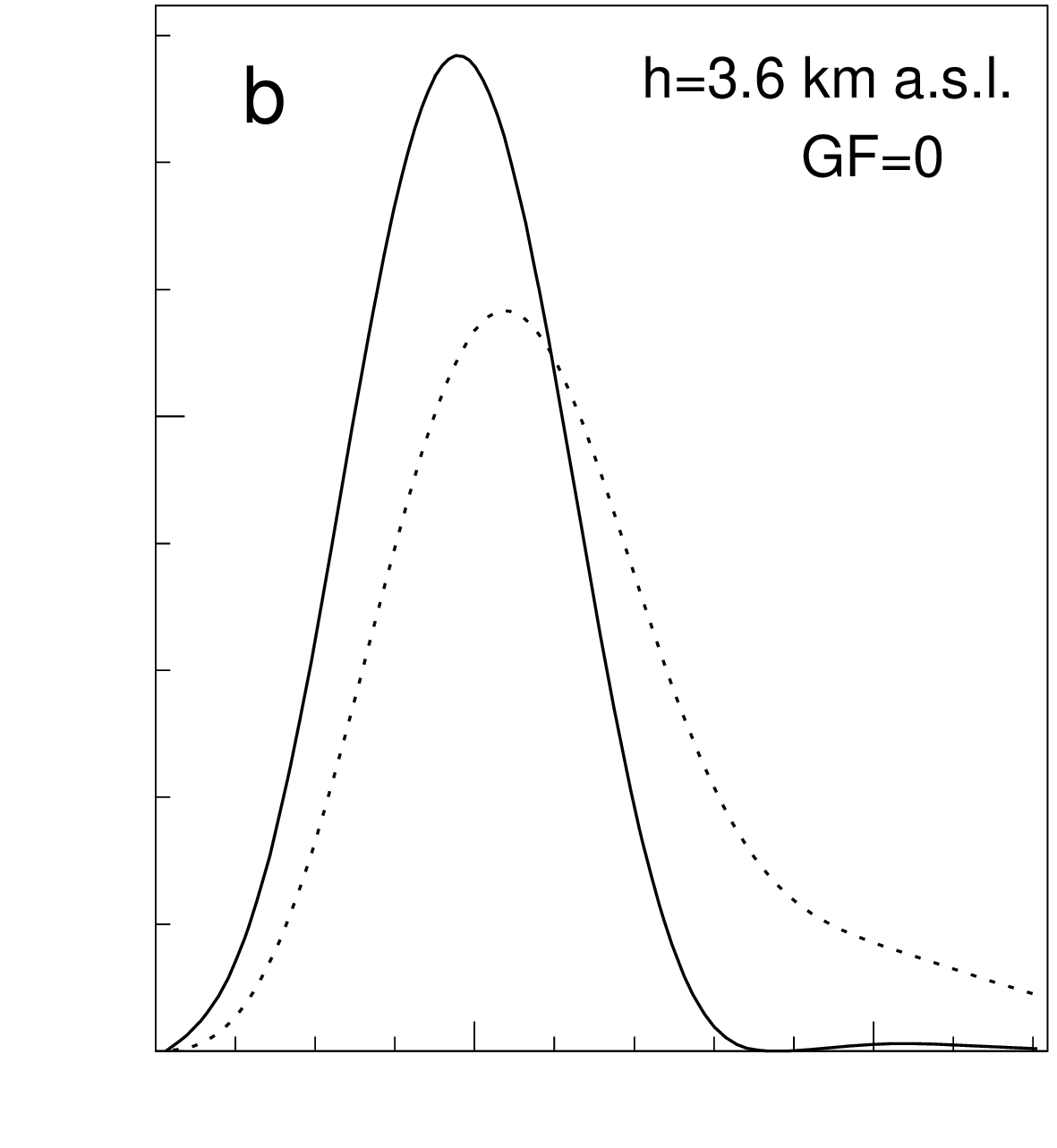}}\\
 \scalebox{0.95}{%
  \includegraphics[width=4.6cm,height=4.6cm]{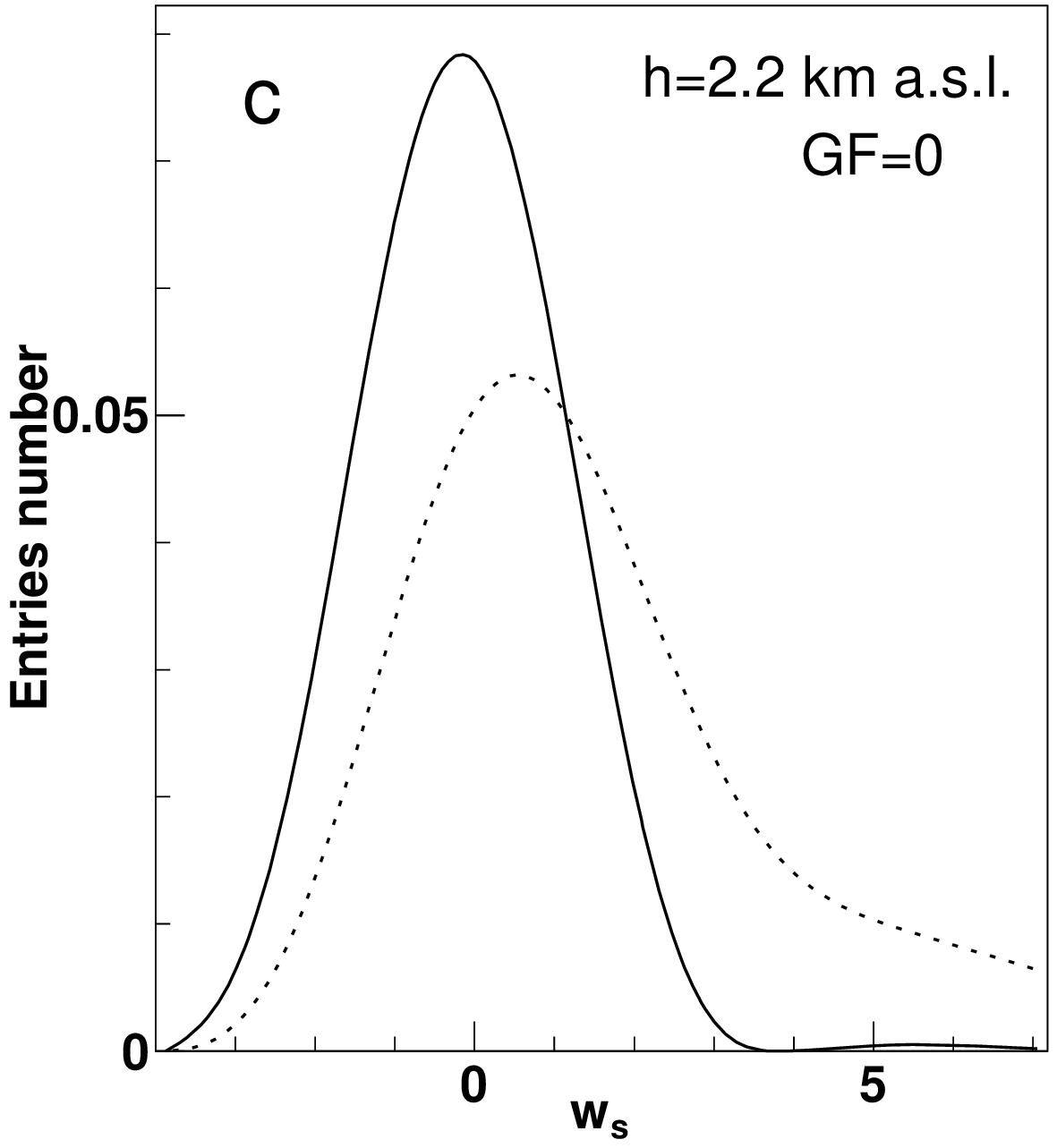}}%
 \scalebox{0.95}{%
  \includegraphics[width=4.6cm,height=4.6cm]{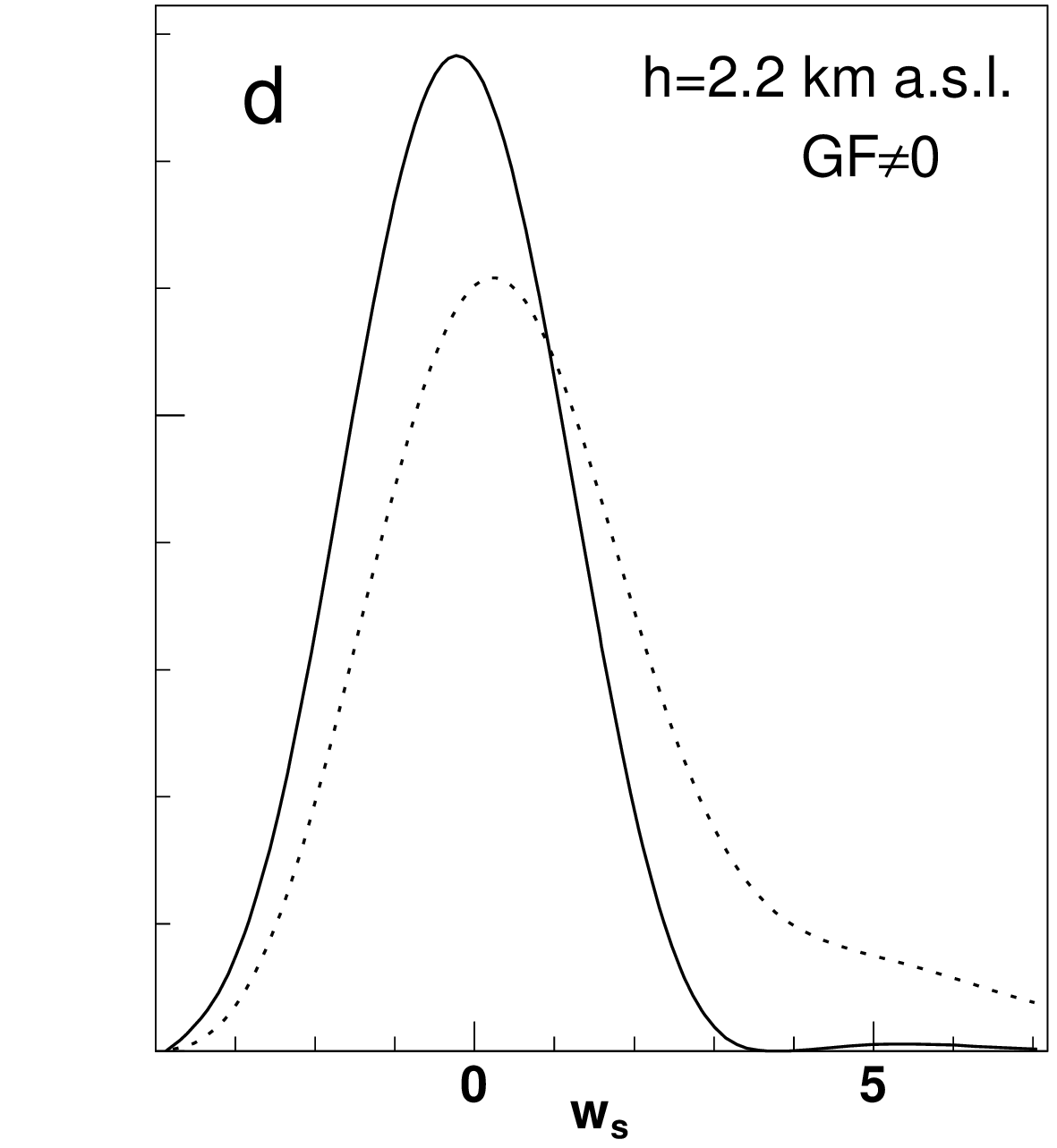}}%
\caption{The scaled width distributions summed up over size bins, for gamma rays (the solid lines) and for protonic background (the dotted lines) at (a) Namibia without GF, (b) Salta without GF, (c) Tenerife without GF and (d) Tenerife with GF at $\phi = 180^{\circ}$. In each panel, the scaled distributions are obtained using the $<\!w\!>$ and $\sigma$ of the gamma ray distribution for the specific altitude and GF. Only events with $Size > 150\,$ pe are used.}
\label{fig:scaled}
\end{center}
\end{figure}

The broadening of gamma ray images, discussed in Section \ref{sec:images}, makes them more similar to hadron images, which are negligibly affected by both the GF and altitude.
Fig.~\ref{fig:scaled} illustrates the related effects in $w_{\rm s}$ distributions. For gamma rays, it always approaches a Gaussian distribution with $<\!w_{\rm s}\!>=0$ and $\sigma=1$. Scaling the protonic {\it width} distribution with $<\!w\!>$ and $\sigma_{\rm width}$ of the broadened (by GF and/or altitude) gamma rays distribution, we get distributions of protonic $w_{\rm s}$ only weakly differing from that of gamma rays. This, obviously, results in a smaller fraction of rejected proton events.

\begin{figure}[t]
\begin{center}
\includegraphics[trim = 8mm 3mm 0mm 2mm, clip,scale=0.44]{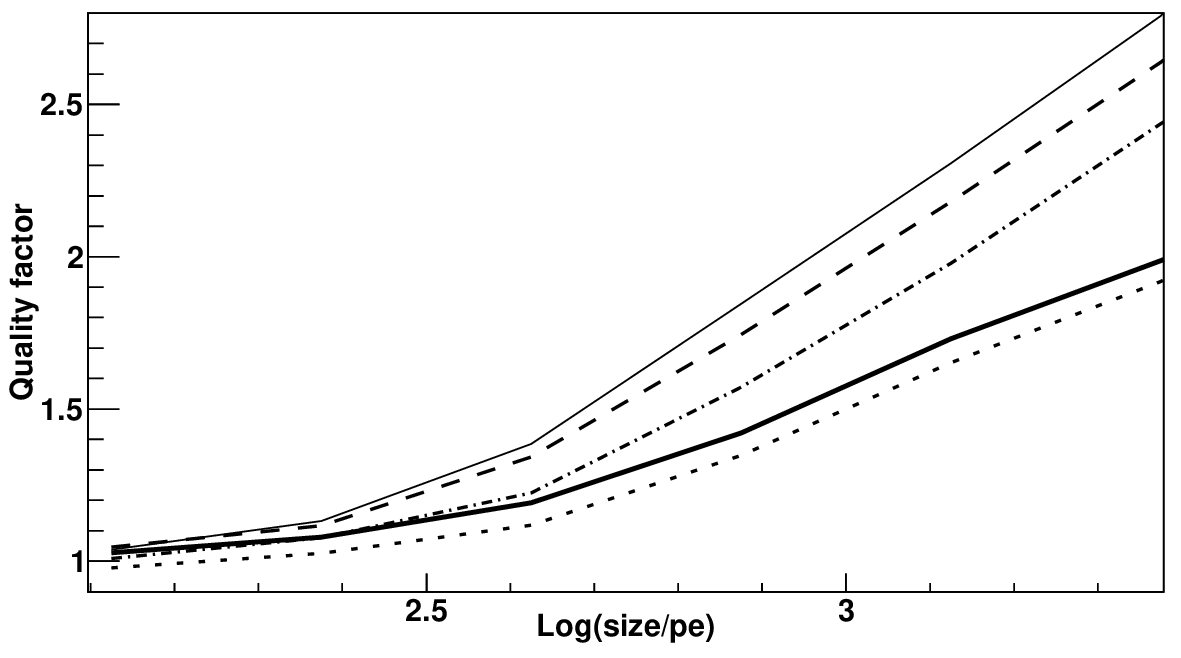}
\caption{The quality factor for the scaled width cut as a function of the size for Namibia (thin-solid), Tenerife (dashed) and Salta (thicker solid) sites without GF and for Tenerife at $\phi = 0^{\circ}$ (dot-dashed) and $\phi = 180^{\circ}$ (dashed) with GF.}
\label{fig:qf}
\end{center}
\end{figure}

We quantify the related effects assuming a primary spectrum with $\Gamma=2.6$. Fig.~\ref{fig:qf} shows the values of the quality factor for the $w_{\rm s}$ cut, with $w_{\rm s}<1$. At small sizes, the separation quality is poor  and  differences between different altitudes or GF strengths are small. However, the differences increase with increasing size (and quality factor); we note that both the increase of the altitude from 1.8 to 3.6 km a.s.l. and the increase of $B_{\perp}$ from 0 to $40 \;{\rm \mu T}$ results in a similar reduction of the QF--as could be expected from magnitudes of effects illustrated in Fig.~\ref{fig:ws}.

Then, we consider a direction cut, for which we accept only events with $\mathit{\Theta}^2 < 0.05$ deg$^2$. We get a much better separation here, with $QF > 5$ even for small sizes. As pointed out in \citep{Commichau08}, the orientation discrimination of $\gamma$-rays  can be significantly degraded by the GF. Indeed, we note a strong reduction of the quality factor for the $\mathit{\Theta}^2$ cut, e.g.\ from $QF=8.4$ (without GF) to $QF = 6.9$ (for $B_{\perp}=40 \;{\rm \mu T}$) for $size = 1000$ pe at Tenerife. We note also that the reduction of the quality factor due to increase of $B_{\perp}$ from 0 to 40 ${\rm \mu T}$ is, independently of size, larger by a factor of $\approx 2.5$ than the reduction due to the increase of altitude from 1.8 to 3.6 km a.s.l., in an approximate agreement with the magnitude of effects illustrated in the left panel of Fig.~\ref{fig:rec}.\\
\indent The increasing GF results in the reduction of both the separation quality and the gamma ray trigger rates, therefore, one can expect that an overall performance should be much worse in a stronger magnetic field. In turn, the altitude effects act in opposite directions at the trigger and analysis levels, so its effects may be expected to partially compensate each other. The numbers of both hadronic and gamma ray events simulated in this work are too small, by at least an order of magnitude, to perform a full analysis and determine the post-analysis parameters, like the energy-dependent sensitivity or effective area. To roughly measure the performance including effects at both levels, we define the effective quality factor similarly as in Eq.~\eqref{eq:qf}, but using the number of simulated gamma ray photons and hadrons, $n^{\gamma,h}_{\rm sim}$, instead of the number of detected events, i.e.\
\begin{align}
QF_{\rm eff}\equiv \frac{ n^{\gamma}_{\rm cuts}/n^{\gamma}_{\rm sim}}{\sqrt{n^{h}_{\rm cuts}/n^{h}_{\rm sim}}}.
\label{eq:qfeff}
\end{align}
While the face value of this parameter depends on the simulation
details (e.g.\ the maximum assumed impact parameters for gamma rays and
protons), its relative changes track both the trigger and separation
efficiencies; note also that it can be used to estimate the minimum detectable 
flux of gamma rays (cf.\ equation (1) in \citep{Aharonian92}).

We find that the effective quality factor decreases from $QF_{\rm eff} = 3.9$ to 1.9 with the increase of GF by 40 ${\rm \mu T}$ at Tenerife, whereas the increase of altitude from 2.2 (Tenerife) 3.6 km a.s.l. (Salta) results in the increase of  $QF_{\rm eff}$ from 3.9 to only 4.3. This illustrates that the GF effect indeed strengthens with respect to the altitude effect in the post-analysis performance. The above changes of $QF_{\rm eff}$ by 50\% (GF) and 10\% (altitude)  indicate that the change of the altitude by $1.4 {\rm km}$ can have a similar effect on the overall performance as changes of the magnetic field by $\sim 10 {\rm \mu T}$, which is the typical difference between the average GF strengths at various sites.

\begin{figure}[t]
\begin{center}
\includegraphics[trim = 8mm 3mm 0mm 2mm, clip,scale=0.44]{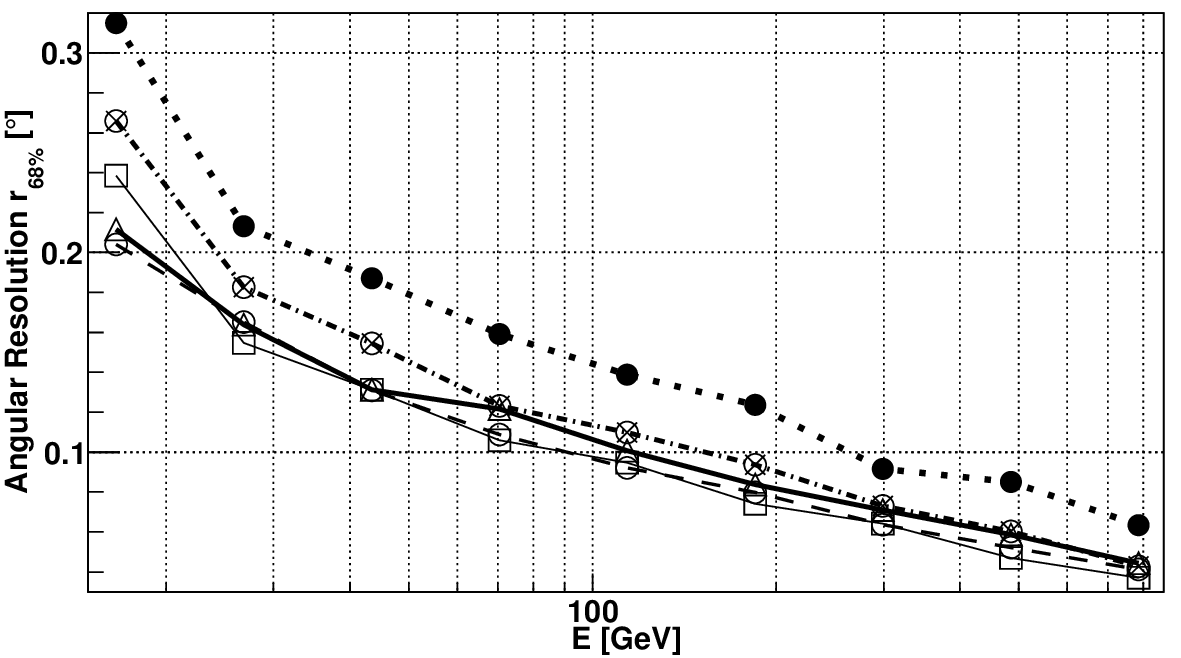}
\caption{Angular resolution as a function of energy for the 68\% containment radius. Three sites are presented: Argentina-Salta without GF (thick-solid line with triangles), Namibia without GF (thinner solid line with boxes), Tenerife without GF (dashed line with empty circles), Tenerife with GF at $\phi = 0^{\circ}$ (dot-dashed line with starred circles) and Tenerife with GF at  $\phi = 180^{\circ}$ (dotted line with filled circles). The energy-dependent cuts in {\it Size} are used (e.g.~85, 300 and 2000 pe (per telescope) for 20, 300 and 500 GeV, respectively).}
\label{fig:ang_res}
\end{center}
\end{figure}

For a primary spectrum with $\Gamma=3.5$ we find similar trends, namely $QF_{\rm eff}$ decreases from 0.40 to 0.16 with the increase of GF by $40\;{\rm \mu T}$ at Tenerife whereas the increase of altitude by $1.4\; {\rm km}$ from Tenerife to Salta results in the increase of $QF_{\rm eff}$ from 0.40 to 0.58, i.e.\ the GF effect is stronger than the altitude effect regardless of the slope of the primary spectrum. $QF_{\rm eff}$ is much smaller for $\Gamma=3.5$ due to larger relative contribution of lower energy events, which are less efficiently rejected, as compared to $\Gamma=2.6$.

We emphasize that our analysis using $QF_{\rm eff}$ is very approximate, in particular neglecting the dependence on photon energy, impact parameter or size as well as neglecting information from other Hillas parameters or, e.g., an additional information from time gradient  \citep{konopelko99}, which could possibly improve the analysis-level performance. Nevertheless, we took into account the most efficient discrimination parameters and the above results indicate that  the GF can {\it strongly} reduce the overall CTA performance if appropriate correction procedures are not developed.

\subsection{Angular resolution}
\label{sec:angular_res}
The angular resolution $\mathit{\Theta_{r}}$, is defined by the so called containment radius, $r_{f}$, containing the fraction, $f$, of events in the distribution of the reconstructed event directions,
i.e.\ it can be determined by the $F(\mathit{\Theta}_{r})=f$ condition, where $F(\mathit{\Theta})$ is the cumulative $\mathit{\Theta}$ distribution (see right panel in Fig.~\ref{fig:rec}).
In Fig.~\ref{fig:ang_res} we show the angular resolution calculated for $f=0.68$ with energy-dependent cuts in {\it Size} (e.g.~85, 300 and 2000 pe (per telescope) for 20, 300 and 500 GeV, respectively).
The increase of $B_{\perp}$ by 40 ${\rm \mu T}$ results in a very strong degradation of the angular resolution, by a factor of three stronger than  the degradation resulting from the increase of the altitude by 1.8 km. 
The influence of the geophysical parameters on the angular resolution is slightly weaker at higher energies, however, the change is small. E.g.\ above the primary energy of 200 GeV the effects are only by $\approx 10\%$ weaker than in a lowest energy regime.

\section{Comparison of sites}
\label{sec:comp}

We first compare the trigger-level performance, for which we made the detailed simulations and which allows to compare parameters directly related to the geophysical condition, without an additional dependence on the analysis procedure.
The magnitude of the GF effect depends on the pointing direction of a telescope, therefore, for each site the performance parameters will change from one source to another. 
For a rough comparison of the sites, we determine the mean values, $R_{\rm tot,av}$ and $E_{\rm th,av}$ (given  in rows 'Av' in Table \ref{tab4}), by linear interpolation between the extreme values (corresponding to  $\phi=0^{\circ}$ and  $\phi=180^{\circ}$) to the average value of the transverse GF, i.e.\ $\left<B_{\perp}(30^{\circ})\right>$.
Thus derived $R_{\rm tot,av}$ and $E_{\rm th,av}$ correspond to the values of the magnetic field indicated by the vertical line in Fig.\ \ref{fig:b}b; as noted in Section 2, they give a good representation of the average GF effect at the considered sites.

The southern sites are located at different altitudes with a similar, moderate GF and, thus, the IACT performance is primarily a function of $h$ at these sites. Salta, located at the largest $h$, has $E_{\rm th,av}$ smaller by almost $\approx 7$ GeV, and $R_{\rm tot,av}$ larger by $\sim 20\%$, than the Namibian site located at the lowest altitude.
 The two northern sites are located at similar altitudes and have similar $\left<B_{\perp}\right>$, as a result, differences between their $R_{\rm tot,av}$ and $E_{\rm th,av}$ do not exceed 5\%.
Note also that the mean performance parameters in the northern sites are similar to those in Namibia.
 
 We emphasize, again,  that parameters characterising a specific direction of observation may significantly deviate from the mean values. The magnitude of these deviations is indicated by the extreme values of performance parameters given in rows 'Min.' and 'Max.' in Table \ref{tab4}. The largest deviations occur for the Mexican site, due to rather large GF strength and its inclination, as noted in Section 2.

Comparing the extreme values of $R_{\rm tot}$ for the Mexican site with $R_{\rm tot,av}$ for Namibia and Salta (see  Table \ref{tab4}), we find that the change of $B_{\perp}$ from 0 to $40 \;{\rm \mu T}$ results in the decrease of the detection rate from 0.210 to 0.122 phot/s, and it is larger by $\approx50\%$ than the decrease, from 0.198 to 0.158 phot/s, corresponding to the change of the altitude from 3.6 to 1.8 km a.s.l.. On the other hand, comparing changes of the threshold energy corresponding to such changes of $B_{\perp}$, from 16.7 to 21.7 GeV, and $h$, from 14.4 to 21.0 GeV, we find that the change of the latter results in a slightly larger, by $\approx 2$ GeV, increase of the threshold energy.

Note that in the particular case of observations along the field lines, the Mexican site are characterised by $R_{\rm tot,min}=0.210\; {\rm phot/s}$ and $E_{\rm th}=16.7\; {\rm GeV}$ similar to $R_{\rm tot,av}=0.198\; {\rm phot/s}$ and $E_{\rm th,av}=14.4\; {\rm GeV}$ at Salta, indicating that even a moderate GF at the latter site has a noticeable effect, compensating the difference of altitudes of $\Delta h = 1.2$ km between these two sites.

\begin{table}[t]
\begin{center}
 \begin{tabular}{|c||c|@{}c@{}|c|c|c|}
  \hline
  \multicolumn{6}{|c|}{\bf Parameters summary}\\
  \hline
    & \multicolumn{2}{|c|}{\bf Northern hemisphere} & \multicolumn{3}{|c|}{\bf Southern hemisphere}\\
  \cline{2-6}
  Site & $\;\;${\bf M}$\;\;$ & {\bf S} & {\bf A-S} & {\bf A-L} & {\bf N}\\
  \hline
  \multicolumn{6}{|@{}c@{}|}{${R_{\rm tot}}$ [phot/s] ($\Delta {R_{\rm tot}}\approx \pm 0.001$ [phot/s])}\\
  \hline
  Min. & $\;\;\;$0.210$\;\;\;$ & 0.172 & 0.214 & 0.190 & 0.187\\
  \hline
  Av. & 0.153 & 0.141 & 0.198 & 0.170 & 0.158\\
  \hline
  Max. & 0.122 & 0.123 & 0.190 & 0.161 & 0.141\\
  \hline
  \multicolumn{6}{|@{}c@{}|}{ $E_{\rm th}$ [GeV] ($\Delta E_{\rm th}= \pm 0.5$ [GeV] )}\\
  \hline
  Min. & 16.7 & 18.7 & 13.7 & 16.9 & 19.2\\
  \hline
  Av. & 20.0 & 21.1 & 14.4 & 17.6 & 21.0\\
  \hline
  Max. & 21.7 & 22.4 & 14.7 & 18.0 & 22.5\\
  \hline
 \end{tabular}
\end{center}
\caption{Gamma trigger rates and the energy threshold for the lowest (Min.), average (specifically $\left<B_{\perp}(30^{\circ})\right>$; Av.) and the highest (Max.) values of $B_{\perp}$ at $\theta=30^{\circ}$ for Argentina-Salta ({\bf A-S}), Argentina-Leoncito ({\bf A-L}), M\`exico-San Pedro Martir ({\bf M}), Spain-Tenerife ({\bf S}) and Namibia-H.E.S.S. ({\bf N}); $\Delta {R_{\rm tot}}$ and $\Delta E_{\rm th}$ are errors for $R_{\rm tot}$ and $E_{\rm th}$, respectively.}
 \label{tab4}
\end{table}

Using the scaling laws derived in Sec ~\ref{sec:results} we estimate in Table \ref{tab5} the energy thresholds and trigger rates for five further sites which are currently considered for the location of CTA observatories. The site in India stands out from other locations due to its very large altitude of 4.5 km. Neglecting the GF effect, we could expect $E_{\rm th} \approx 9$ GeV at this site. However, the GF is strong at this site resulting in $E_{\rm th,av}\approx14$ GeV, similar to that in Salta. Taking into account experiments other than CTA, we note that a similar conclusion applies to the location of LHAASO \citep{Lhaaso}, which is planned in Tibet at a similar altitude (4.3 km) and in a similarly strong GF. We caution, however, that both the altitude and the GF strength at these two sites exceed the maximum values for which we made computations and verified  our linear relations, so the extrapolations come with an obvious risk.

The magnitude of the changes of the performance parameters with varying observation angle is a separate factor for evaluating a site. In general, sites with $H<|Z|$ are characterized by larger changes, at least for small or moderate $\theta$. On the positive side, sites with $H<|Z|$ have a good performance at small observation angles, for which $B_{\perp}$ is roughly equal to the (smaller) $H$ component.
Furthermore, the relative changes of performance are slightly larger at smaller altitudes (an effect corresponding to the decrease of the altitude-dependent factors,  $E_{\rm th}^0(h)$ and  $R_{\rm tot}^0(h)$, in Eqs \ref{eq:E_th} and \ref{eq:R_tot}) and, obviously, in a stronger GF.
Among the considered sites, a less uniform performance can be expected at the northern hemisphere, where the GF is in general stronger. Two sites with $|Z|>H$, in US and Mexico, are characterised by the largest variations, e.g.\ a factor of 2 changes of the detection rates between the southward and northward directions. The third northern site with $|Z|>H$, in India, also having the strongest GF, is located at the largest altitude which partially reduces the performance inhomogeneity (as explained above), e.g.\ the changes of the detection rates to a factor of 1.5.
Tenerife, with $H>|Z|$,  has a more uniform performance at $\theta \le 30^{\circ}$ but  stronger variations can be expected for larger $\theta$, especially approaching $60^{\circ}$.
At African sites $|Z|$ is larger than $H$, however, the GF is weaker here than at the northern sites and the changes are less pronounced, e.g.\  variations of the detection rate do not exceed a factor of $1.5$. 
The  South American sites have $|Z|<H$, rather weak GF and large altitudes; as a result they are characterised by a very homogeneous performance.

Concluding the trigger-level comparison, we note that

\noindent
(1) the mean parameters are similar for most of the ten considered sites, with, in particular,  $E_{\rm th,av}$ around 20 GeV, and the differences of $R_{\rm tot,av}$ and $E_{\rm th,av}$ not exceeding 10 per cent, except for sites discussed in (2) below;

\noindent
(2) the Argentinian (Salta) and Indian sites provide the best perspective for approaching the trigger-level energy threshold of 10 GeV;

\noindent
(3) our scalings laws allow to estimate the amount of the performance changes for given values of $H$, $Z$ and $h$;
the largest performance changes characterise the US and Mexican sites, while the weakest changes occur at the South American sites.

For the analysis level, we cannot discuss the energy dependent effects, in particular, the differences in the post-analysis energy threshold. However, the differences between the quality factors assessed in Section 4.4 give some hints for trends in the post-analysis performance. 
In general, they indicate that our result at the trigger level should be taken into account as constraints limiting the altitude effect from above (i.e.\ the post-analysis performance should be less sensitive to altitude) and the GF effect from below.
The differences of the average $<B_{\perp}>$ among all considered sites do not exceed 15 $\mu T$, and
among sites of each hemisphere considered separately, they do not exceed 5 $\mu T$. Therefore,
at the trigger level the average performance parameters are more significantly affected 
by the altitude than the GF.  When the efficiency of
hadron rejection is taken into account,  even the average GF may be similarly important for the overall performance (very roughly estimated by $QF_{\rm eff}$) as the altitude.
However, at the current stage any further statement would be premature.\\
We note also that the angular resolution (assessed here without applying any separations cuts) may be markedly reduced at sites with stronger GF.

\begin{table}[h]
\begin{center}
 \begin{tabular}{|@{}c@{}|c|c|c|c|c|c|c|}
  \hline
  \multicolumn{3}{|@{}c@{}|}{Site} & { \bf B} & {\bf U-Y} & { \bf N-L} & {\bf C} & {\bf I}\\
  \hline
  \multicolumn{8}{|@{}c@{}|}{Localization}\\
  \hline
  \multicolumn{3}{|@{}c@{}|}{Lat.~[$^{\circ}$]} &-32.2 & 35.1 & -26.6 &-29.3 & 32.8\\
  \multicolumn{3}{|@{}c@{}|}{Lon.~[$^{\circ}$]}& 22.5 &-112.9 & 16.6  &-70.7 & 79.0\\
  \hline
  \multicolumn{3}{|@{}c@{}|}{Height} & 1.6 & 1.7 & 1.76 & 2.4 & 4.5\\
  \multicolumn{3}{|@{}c@{}|}{[km a.s.l.]} &  &  &  &  &\\
  \hline
  \multicolumn{8}{|c|}{Local magnetic field $\vec{B}=(H,0,Z)$ [$\mu$T]} \\
  \hline
  \multicolumn{3}{|@{}c@{}|}{$H$}& 11.0 & 23.6 & 11.0 & 20.6 & 31.7\\
  \multicolumn{3}{|@{}c@{}|}{$Z$ (+$\downarrow$)}& -24.0 & 42.8 & -24.9 & -11,3 & 38.7\\
  \hline
  \multicolumn{8}{|c|}{$B_{\perp}(\theta=30^{\circ})$ [$\mu$T]}\\
  \hline
  \multicolumn{3}{|@{}c@{}|}{$B_{\perp}^{\rm\; min}$} & 2.5 & 1.0 & 2.9 & 12.2 & 8.1\\
  \hline
  \multicolumn{3}{|@{}c@{}|}{$B_{\perp}^{\rm\; av}$} & 14.5 & 28.0 & 14.9 & 19.7 & 33.0\\
  \hline
  \multicolumn{3}{|@{}c@{}|}{$B_{\perp}^{\rm\; max}$} & 21.5 & 41.8 & 22.0 & 23.5 & 46.8\\
  \hline
  \multicolumn{8}{|c|}{$R_{\rm tot}$ [phot/s]}\\
  \hline
  \multicolumn{3}{|@{}c@{}|}{Min.} & 0.18 & 0.18 & 0.18 & 0.18 & 0.26\\
  \hline
  \multicolumn{3}{|@{}c@{}|}{Av.}  & 0.15 & 0.12 & 0.16 & 0.17 & 0.21\\
  \hline
  \multicolumn{3}{|@{}c@{}|}{Max.} & 0.13 & 0.09 & 0.14 & 0.16 & 0.18\\
  \hline
  \multicolumn{8}{|@{}c@{}|}{$E_{\rm th}$ [GeV]}\\
  \hline
  \multicolumn{3}{|@{}c@{}|}{Min.} & 20.1 & 19.5 & 19.5 & 18.6 & 10.4\\
  \hline
  \multicolumn{3}{|@{}c@{}|}{Av.} & 21.9 & 23.5 & 21.4 & 19.8 & 14.2\\
  \hline
  \multicolumn{3}{|@{}c@{}|}{Max.} & 22.9 & 25.6 & 22.5 & 20.3 & 16.2\\
  \hline
 \end{tabular}
\end{center}
\caption{Trigger rates and energy thresholds obtained from  Eq.\ (\ref{eq:R_tot}) and (\ref{eq:E_th}) for South Africa-Beaufort West ({\bf B}), USA-Yavapai Ranch ({\bf U-Y}), Namibia-Lalapanzi ({\bf N-L}), Chile-La Silla ({\bf C}) and India-Hanle ({\bf I}). The values of $H$, $Z$ and $B_{\perp}$ are obtained from the data given by {\it National Geophysical Data Center} at {\it www.ngdc.noaa.gov/geomag}; for each site we take the GF parameters for the altitude of 10 km a.s.l.. Detailed information about particular sites are available in Chapt.~9 of \citep{cta}.}
 \label{tab5}
\end{table}

\section{Summary and discussion}
\label{sec:summary}

We have studied the influence of two geophysical factors, the local GF and altitude, on the low-energy performance of the planned CTA observatories. We quantify the changes of the performance parameters, which may be used to balance the geophysical conditions against other criteria for site selection.
We derive an approximately linear scaling of the trigger-level threshold energy and detection rates with both $B_{\perp}$ and $h$, which is a novel result.

Previous detailed studies of the GF effect were performed only for two sites with rather large magnetic field, namely, the MAGIC-site in La Palma, with the GF similar to that in Tenerife, and Narrabri with even larger GF. The magnitude of the effect on the energy threshold in moderate GF, or its comparison to the altitude effect, remained rather uncertain. Recently a rough estimation of the GF effect in the candidate CTA sites was independently done in Bachelor thesis of Maria Krause \citep{Krause11}. Under slightly different assumptions on configuration of the telescope array 
(with, in particular, 9 large telescopes and 80 m spacing) she derived the difference   in performance parameters between the southern and northern directions of observation approximately consistent with ours.  

Regarding the altitude effect, the possibility of the detection of $\sim 10$ GeV photons at altitudes of $\sim 5$ km was considered already in early simulations \citep{browning77} and more recently it was emphasized in \citep{Aharonian01}. Indeed, assuming that our linear scaling of Eq.~\eqref{eq:E_h}  holds at large altitudes we find that the threshold energy below 10 GeV could be achieved by an observatory, with parameters assumed for CTA, located at altitudes exceeding 4.3 km in the {\it absence of GF}. However, the currently considered sites with such altitudes in India (for CTA) and in Tibet (for LHAASO) have rather strong GF which prevents achieving threshold energies below 10 GeV, except for the specific case of observations along the field lines. 

We made a basic estimation of the distortion of image parameters, which are crucial for the gamma/hadron separation and the direction reconstruction. As previously suggested by \citep{Konopelko04}, and confirmed in our calculations, the gains in trigger efficiency with increasing altitude are significantly reduced at the analysis level, as the extra photons have images which are less suitable for the separation. We confirm also the conclusion of \citep{Commichau08} that the rotation of images by the GF may be the major obstacle in the IACT performance. We find that the maximum changes of the GF and altitude considered here (i.e. 40 $\mu T$ and 1.8 km) have similar (in magnitude) effects in the distortion of width profiles and in the efficiency of the scaled width cuts. On the other hand, the GF effect is much stronger (by a factor of $\sim 3$) in the angular resolution and in the related efficiency of the direction cuts. We notice that while the trigger-level effects are important only at low energies (around the threshold), the effects in the hadron rejection efficiency or in the reconstruction quality are significant at all energies.

We also made a very basic (in particular, energy-independent) estimation of the overall performance, which indicates that even the average GF may be equally important as the altitude of a site. Furthermore, a site with strong variations of $B_{\perp}$ for different pointing directions will be characterised by a very inhomogeneous performance with respect to the observation direction.

Our qualitative conclusions are independent of the slope of the primary spectrum. The threshold energy is less sensitive to changes of both $B_\perp$ and $h$ for a softer spectrum. However, both the GF and the altitude effects on the detection rates appear independent of the primary spectrum slope.

We  performed high-statistics gamma ray simulations for five sites and we applied our scaling laws for five additional sites.
Among  the ten considered sites, only the Argentinian and Indian sites have notably  better, trigger-level performance parameters than the remaining sites. In the post-analysis performance the GF effect will be strengthened and the altitude effect weakened, however, we note again that we regard the trigger-level information as more basic because it does not depend on the assumed analysis procedure.

\section*{Acknowledgments}
\label{sec:acknowledgments}
We thank Tomasz Bulik, Rene Ong and anonymous Reviewers for their remarks.
This work was supported by the NCBiR grant ERA-NET-ASPERA/01/10 and NCN grant UMO-2011/01/M/ST9/01891.

\bibliographystyle{elsarticle-num}
\bibliography{gf_cta}

\begin{thebibliography}{10}
\expandafter\ifx\csname url\endcsname\relax
  \def\url#1{\texttt{#1}}\fi
\expandafter\ifx\csname urlprefix\endcsname\relax\def\urlprefix{URL }\fi
\expandafter\ifx\csname href\endcsname\relax
  \def\href#1#2{#2} \def\path#1{#1}\fi

\bibitem{buckley08}
J.~{Buckley}, et~al.$<$arXiv:0810.0444$>$.

\bibitem{Aleksic11}
J.~Aleksi{\'c}, et~al., Astropart.~Phys. 35 (2012) 435--448.

\bibitem{Hofmann00}
W.~Hofmann, et~al., in: AIP Conf.~Proc., Vol. 515, 2000, pp. 500--509.

\bibitem{Veritas08}
J.~Holder, et~al., in: AIP Conf.~Proc., Vol. 1085, 2008, pp. 657--660.

\bibitem{cta}
M.~Actis, et~al., Exp.~Astron. 32 (2011) 193--316.

\bibitem{Cocconi54}
G.~Cocconi, Phys.\ Rev.\ 93 (1954) 646--647.

\bibitem{browning77}
R.~{Browning}, K.~E. {Turver}, Nuovo Cimento A Serie 38 (1977) 223--238.

\bibitem{Elbert83}
J.~W. Elbert, T.~Stanev, S.~Torii, in: Int.~Cosmic Ray Conf.~Proc., Vol.~6,
  1983, pp. 227--230.

\bibitem{bowden92}
C.~{Bowden}, et~al., J.~Phys.~G 18 (1992) L55--L60.

\bibitem{Commichau07}
S.~C. Commichau, et~al., in: Int.~Cosmic Ray Conf.~Proc., Vol.~3, 2007, pp.
  1357--1360.

\bibitem{Commichau08}
S.~C. Commichau, et~al., Nucl.~Instrum.~Methods Phys.~Res., Sect.~A 595 (2008)
  572--586.

\bibitem{Beisembayev99}
R.~U. Beisembayev, in: Int.~Cosmic Ray Conf.~Proc., Vol.~1, 1999, p. 471.

\bibitem{Chadwick99}
P.~M. Chadwick, et~al., J.~Phys.~G 25 (1999) 1223--1233.

\bibitem{PAO11}
P.~Abreu, et~al., J.~Cosmology Astropart.~Phys. 11 (2011) 022.

\bibitem{Aharonian01}
F.~A. Aharonian, et~al., Astropart.~Phys. 15 (2001) 335--356.

\bibitem{Campbell03}
W.~H. Campbell, Introduction to Geomagnetic Fields, Cambridge Univ.~Press,
  2003.

\bibitem{Firpo06}
R.~Firpo, Master's thesis, Universitat Aut\`onoma de Barcelona (2006).

\bibitem{Aharonian06}
F.~Aharonian, et~al., Astronom.~Astrophys. 457 (2006) 899--915.

\bibitem{Heck98}
D.~Heck, et~al., Tech.~Rep.~FZKA 6019.

\bibitem{Bernlohr08}
K.~Bernloehr, Astropart.~Phys. 30 (2008) 149--158.

\bibitem{Bernlohr00}
K.~Bernloehr, Astropart.~Phys. 12 (2000) 255--268.

\bibitem{Kneizys96}
F.~X. Kneizys, et~al., The modtran 2/3 report and lowtran 7 model, Tech. rep.,
  Phillips Laboratory, Hanscom AFB, MA (USA), 01731--3010 (1996).

\bibitem{BernlohrCTA08}
K.~Bernloehr, in: AIP Conf.~Proc., Vol. 1085, 2008, pp. 874--877.

\bibitem{Bernlohr12}
K.~Bernloehr, et~al., Monte carlo design studies for the cherenkov telescope
  array, Astropart.~Phys.(in preparation),$<$arXiv:1210.3503$>$.

\bibitem{Daum97}
A.~Daum, et~al., Astropart.~Phys. 8 (1997) 1--11.

\bibitem{hinton09}
J.~A. {Hinton}, W.~{Hofmann}, Ann.~Rev.~Astronom.~Astrophys. 47 (2009)
  523--565.

\bibitem{Konopelko04}
A.~Konopelko, J.~Phys.~G 30 (2004) 1835--1846.

\bibitem{albert08b}
J.~Albert, et~al., Nucl.~Instrum.~Methods Phys.~Res.~A 588 (2008) 424--432.

\bibitem{konopelko99}
A.~Konopelko, et~al., Astropart.~Phys. 10 (1999) 275--289.

\bibitem{Hillas85}
A.~M. Hillas, in: Int.~Cosmic-Ray Conf.~(La Jolla) Proc., Vol.~3, 1985, pp.
  445--448.

\bibitem{Fegan97}
D.~J. Fegan, J.~Phys.~G 23 (1997) 1013--1060.

\bibitem{Sobczynska09}
D.~Sobczy\'nska, J.~Phys.~G 36 (2009) 045201.

\bibitem{Aharonian92}
F.~A. Aharonian, et~al., Exp.~Astr. 2 (1993) 331--344.

\bibitem{Lhaaso}
C.~Zhen, Chin.~Phys. C 34 (2010) 249--252.

\bibitem{Krause11}
M.~Krause, Master's thesis, University of Cottbus (2011).

\end{thebibliography}

\end{document}